\documentclass[a4paper,11pt]{article}
\pdfoutput=1 

\usepackage{jheppub} 

\usepackage[T1]{fontenc} 
\usepackage{units}
\usepackage{subcaption}
\usepackage{graphicx}
\usepackage{float}
\RequirePackage[dvipsnames]{xcolor}

\newcommand{\Gq}[6]{\mathcal{G}^{#1}(#2,\boldsymbol{#3}; #4, \boldsymbol{#5}| #6)}
\newcommand{\Gaq}[6]{\overline{\mathcal{G}}^{#1}(#2,\boldsymbol{#3}; #4, \boldsymbol{#5}| #6)}
\newcommand{\Gg}[6]{\mathcal{G}^{#1}_A(#2,\boldsymbol{#3}; #4, \boldsymbol{#5}| #6)}

\newcommand{\Gqd}[6]{(\mathcal{G}^{\dagger})^{#1}(#2,\boldsymbol{#3}; #4, \boldsymbol{#5}| #6)}
\newcommand{\Gaqd}[6]{(\overline{\mathcal{G}}^{\dagger})^{#1}(#2,\boldsymbol{#3}; #4, \boldsymbol{#5}| #6)}
\newcommand{\Ggd}[6]{(\mathcal{G}^{\dagger})^{#1}_A(#2,\boldsymbol{#3}; #4, \boldsymbol{#5}| #6)}

\newcommand{\Exp}[1]{\exp{\left\{#1\right\}}}
\title{\boldmath  Gluon to $q\bar q$ antenna in anisotropic QCD matter: spin-polarized and azimuthal jet observables}

\newcommand{\nn}{\nonumber\\ }

\def\P{{\boldsymbol P}}

\def\p{{\boldsymbol p}}

\def\x{{\boldsymbol x}}
\def\y{{\boldsymbol y}}
\def\z{{\boldsymbol z}}

\def\r{{\boldsymbol r}}

\def\Re{\text{Re} \, }

\newcommand{\cK}{\mathcal{K}}

\newcommand{\cA}{\mathcal{A}}

\author[a]{Jo\~ao Barata,}
\author[b,c]{Carlos A. Salgado,}
\author[b,d,e]{Jo\~ao M. Silva}
\affiliation[a]{Physics Department, Brookhaven National Laboratory, Upton, NY 11973, USA}
\affiliation[b]{Instituto Galego de Física de Altas Enerxías IGFAE, Universidade de Santiago de Compostela, E- 15782 Galicia-Spain}
\affiliation[c]{Axencia Galega de Innovación (GAIN), Xunta de Galicia, Galicia-Spain}
\affiliation[d]{Laboratório de Instrumentação e Física Experimental de Partículas (LIP), Av. Prof. Gama Pinto, 2, 1649-003 Lisbon, Portugal}
\affiliation[e]{Departamento de Física, Instituto Superior Técnico (IST), Universidade de Lisboa, Av. Rovisco Pais 1, 1049-001 Lisbon, Portugal}

\emailAdd{jlourenco@bnl.gov,joao.m.da.silva@tecnico.ulisboa.pt,carlos.salgado@usc.es }

\abstract{We study the production of a quark-antiquark antenna in the presence of a dense and anisotropic QCD medium. We assume the antenna to originate from an unpolarized gluon state, and consider both massless and massive final states. The medium anisotropy is captured by allowing the jet quenching coefficient to take different magnitudes in orthogonal directions with respect to the jet axis. We find that the final particle distribution is sensitive to the medium anisotropy, and more importantly, that this effect couples directly to the helicity/spin of the final states. We propose to look into these effects by performing a Fourier decomposition of the particle distribution inside the jet. In our medium model, we find that the spin independent terms contribute to the even harmonics of the cosine series. The helicity/spin dependence enters only through the sine Fourier series. We further explore the spin dependence by examining the degree of polarization of the final states in different directions. Our results indicate that the anisotropies present in the QCD matter produced in heavy ion collisions can be probed by studying azimuthal and spin observables inside jets.
}

\begin{document} 
\maketitle
\flushbottom

\section{Introduction}
\label{sec:intro}

The quark-gluon plasma (QGP) is a hot and dense QCD state of matter produced in the early epochs of the Universe, where quarks and gluons are not confined inside hadrons. On Earth, the QGP has been experimentally produced in the aftermath of ultrarelativistic heavy ion collisions (URHICs), at the Relativistic Heavy Ion Collider (RHIC)\cite{PHENIX:2001hpc, STAR:2005gfr, STAR:2020xiv} and the Large Hadron Collider (LHC) \cite{Muller:2012zq,ATLAS:2018gwx, CMS:2021vui}. These experiments have allowed to test the many-body properties of QCD in the most extreme regimes, and lead to the remarkable discovery that the QGP is a nearly ideal liquid, see e.g.~\cite{Busza:2018rrf} for a review on URHICs.

The QGP produced at RHIC and the LHC appears as an intermediate phase of a complex evolution, bridging a sequence of different QCD states of matter. Generally, it is believed that in the first $1$ fm/c (or $3$ ys) after the ions collide, the produced matter is in a highly anisotropic and out-of-equilibrium state, dominated by strong polarized chromoelectric and magnetic fields, which eventually evolve into an equilibrium state preceding the QGP phase~\cite{Schlichting:2019abc,Berges:2020fwq}. Afterwards, the QGP expands hydrodynamicaly, with spatial gradients driving the evolution, until the plasma temperature is sufficiently low, at which point the confinement mechanism leads to the production of a gas of hadrons, later detected by the experiment~\cite{Jaiswal:2016hex,Busza:2018rrf}.\footnote{Despite the existence of several sophisticated models describing the many parts of this evolution, there is no overall widely accepted theoretical picture, nor are the current measurements capable of probing all the aspects characterizing the different stages of URHICs uniquely.} To probe the QCD matter produced in URHICs, and ultimately extract the properties of the QGP, it is common to use energetic probes generated in the same event, which are sensitive to the underlying medium as they evolve through it. One of the most popular and powerful ones are hadronic jets, i.e. collimated QCD cascades originating from highly boosted partons, whose fragmentation occurs simultaneously with the matter evolution. The presence of a background medium induces modifications to the final particle distribution inside jets, with part of the energy thermalizing to the medium. The study of these medium induced effects on jets is typically referred to as jet quenching~\cite{Qin:2015srf, Cunqueiro:2021wls, Apolinario:2022vzg}. Since jets' evolution covers a wide range of energy scales, they offer an ideal probe of the spacetime structure of the QGP and its transport properties.

So far, most theoretical jet quenching studies have focused on understanding how mechanisms such as parton momentum broadening or stimulated radiative energy loss affect jet observables and can be related to fundamental properties of the plasma. Such studies are typically performed using simplified medium models, mainly focused on capturing the QGP phase, commonly assuming the medium to be static, isotropic, and homogeneous. However, as mentioned above, the states of matter produced in real URHICs can be highly dynamical and anisotropic, either due to the presence of polarized fields or due to spatial gradients, for example. If such type of effects are sufficiently strong, one should expect the final particle distribution inside jets to reflect an angular \textit{modulation}, which can be correlated to the structure of the medium.\footnote{We note nonetheless that since the hydrodynamic expansion tends to isotropize the system, many of these effects can be ultimately washed out.} The observation of such an effect is highly non-trivial from the experimental point of view, since it requires correlating the jet measurement to soft sector physics, but it would allow to use jets as detailed probes of the structure of matter in URHICs. From the theory side, including such contributions into account is technically challenging, and requires further expanding the current theoretical toolkit~\cite{Wang:2002ri,Vitev:2004bh,Armesto:2004pt}.

In this work, we take a step in this direction, by studying how a quark-antiquark state branching from an initial unpolarized gluon couples to the anisotropies of an underlying QCD medium, similarly to what was done recently for pure gluonic systems~\cite{Hauksson:2023tze}. Assuming the medium to be sufficiently long, we compute the leading order cross-sections for this process in the large $N_c$ limit, and show that the final distributions are sensitive to the matter structure. In particular, we find that helicity/spin states couple distinctly to the medium anisotropies, and thus they offer a way to extract the details of the underlying matter, not accessible with standard jet observables. To this end, we propose extracting the azimuthal harmonic inside jets, similar to what is done in the soft sector of URHICs, see also~\cite{Baty:2021ugw,Zhao:2024wqs}. We show below that, for the case where the jet transport coefficient is distinct in two orthogonal directions transverse to the jet axis, the leading terms in the harmonic decomposition of the particle distribution inside a jet, for small anisotropy, read
\begin{align}
   \frac{2\pi}{ dN^h/dz} \frac{dN^h}{dzd\phi}  = 1+ v_2 \cos\left(2 \phi \right) + w_2^{(h)} \sin\left(2 \phi \right)\, , 
\end{align}
where $dN^h$ denotes the helicity/spin dependent final particle distribution, $\phi$ is the azimuthal angle around the jet axis, $z$ is the energy sharing fraction of the antenna, $v_2,w_2$ vanish in isotropic matter, and all the helicity/spin dependence lies in $w^{(h)}_{2}$.

This work is structured as follows. In Section~\ref{sec:calculation}, we provide the $g \to q\bar q$ cross-section in the presence of an anisotropic medium, for both massless and massive final states. We discuss jet observables capable of probing and extracting the features of the medium anisotropy in Section~\ref{sec:observables}, followed by a summary and discussion of our results in Section~\ref{sec:conclusion}. Several technical details of the calculation are shown in Appendices~\ref{App1} and~\ref{app:App2}.

\section{Quark pair production in a dense anisotropic QCD medium }
\label{sec:calculation}

\begin{figure}
    \centering
    \includegraphics[width=.75\textwidth]{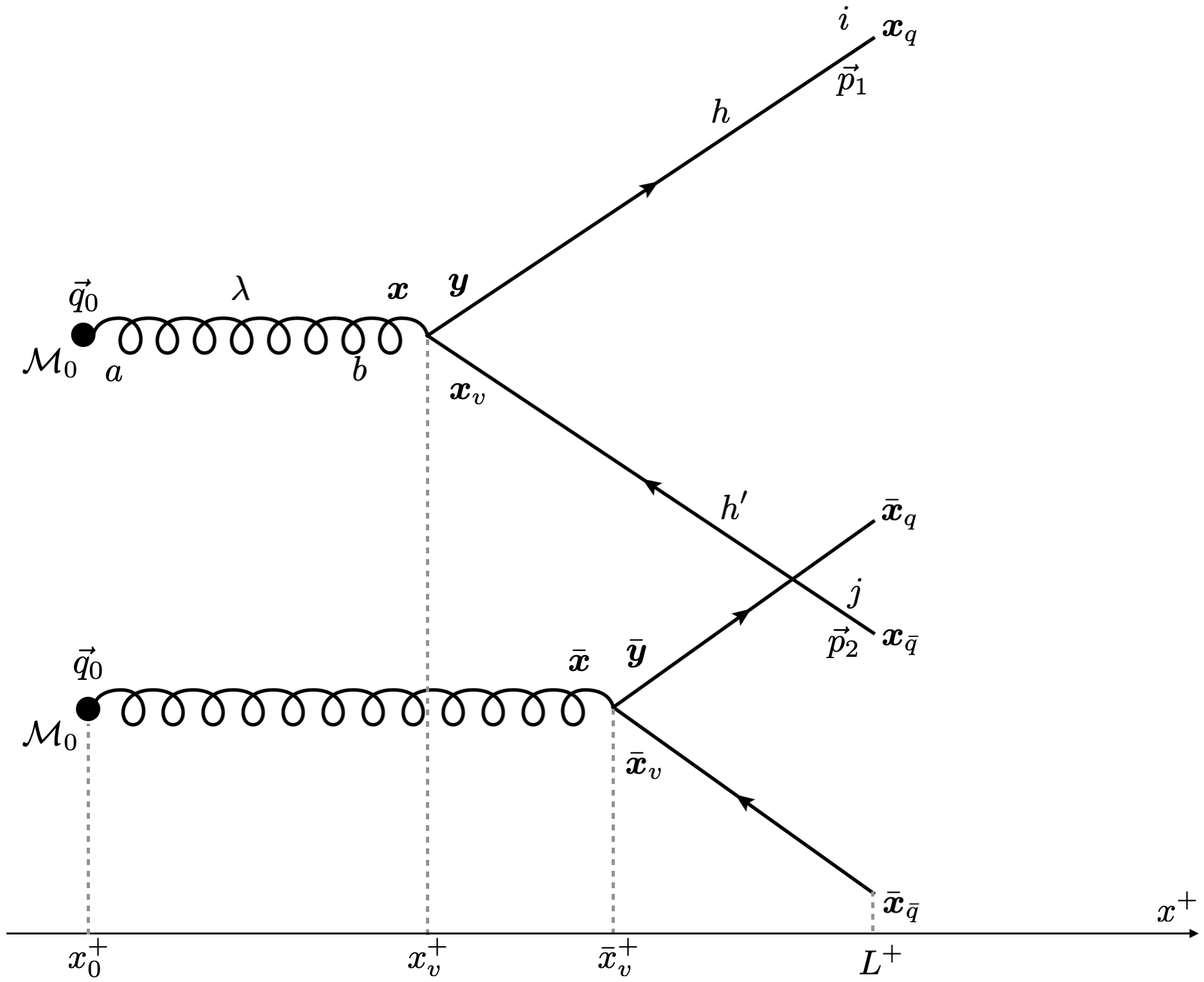}
    \caption{Leading order diagrams contributing to the antenna production. We indicate the most relevant spacetime, polarization and color indices entering the main text calculation, representing the (conjugate) amplitude on top (bottom). The production point and initial momentum of the gluon is assumed to be the same on both sides of the cut. Notice that although the diagram is the same as in vacuum propagation, here one should associate a medium modified propagator to each partonic leg. }
    \label{fig:diagram}
\end{figure}

We start by deriving the double differential cross-section for the process presented in Fig.~\ref{fig:diagram}. This consists in the branching of a gluon to a $q\bar q$ pair (i.e. a quark antenna) inside a medium of length $L^+$ and density $n(x^+)$, where $x^+$ denotes the light-cone time. 
We shall assume that the medium length is sufficiently long, such that we only take into consideration processes where the branching takes place inside the medium. Nonetheless, the calculation shown below can be easily extended to fragmentation in finite media. We compute the quark antenna cross-section for both massless and massive final states.

The most important element in our calculation is that we assume the medium to be anisotropic. In particular, following~\cite{Hauksson:2023tze}, we incorporate this picture by allowing the jet quenching parameter $\hat q$ to take different values along two orthogonal directions in the transverse plane to the jet axis, which we shall denote as $x$ and $y$. The case when $\hat q_x=\hat q_y$ corresponds to the limit of an isotropic background. As we show below, even when considering an unpolarized initial state, the final particle distribution has a non-trivial azimuthal profile which depends on the final states' polarization. This suggests that studying the helicity/spin dependent final state distributions inside jets can give non-trivial information regarding the structure of the  underlying medium; we further explore this idea in the next section. Nonetheless, we note that our matter model is rather simplistic, and many effects that tend to isotropize the system, could qualitative change the results. We will disregard such contributions in this work. Finally, in what follows we shall not discuss the physics from which the anisotropic $\hat q$ emerges. We note that equivalent forms for this parameter have been found in the early stages of heavy ions collisions~\cite{Boguslavski:2023waw,Boguslavski:2023alu,Ipp:2020mjc,Ipp:2020nfu,Avramescu:2023qvv,Carrington:2021dvw,Carrington:2022bnv,Barata:2024xwy}, and in the QGP epoch~\cite{Kuzmin:2023hko,Barata:2023zqg,Barata:2023qds,Barata:2022utc,Andres:2022ndd,Barata:2022krd,Sadofyev:2021ohn,He:2020iow,Xiao:2024ffk,Fu:2022idl,He:2022evt,Kuzmin:2024smy}.\footnote{See also~\cite{Hauksson:2020wsm,Hauksson:2021okc} for related work.} How all these different sources give rise to an anisotropic transport coefficient and its impact on the evolution of the partonic cascade in the medium are active topics of research. The observables proposed in this work would contribute to pin down the relevant dynamics associated to an anisotropic transport coefficient.

We consider first the case of massless quark states. Since we are interested in the limit of a dense QCD background medium, when computing the diagrams shown in Fig.~\ref{fig:diagram}, one has to resum contributions describing the scattering of the jet partons with the medium constituents. This can be efficiently performed in a path integral formalism, as originally discussed by BDMPS-Z~\cite{Zakharov:1996fv,Baier:1996kr}, see also~\cite{Attems:2022ubu} for a recent related calculation. Very briefly,\footnote{See e.g.~\cite{Mehtar-Tani:2013pia,Blaizot:2015lma} for recent reviews.} in this approach, to each line in Fig.~\ref{fig:diagram} one associates an effective quantum mechanical propagator, which for massless quarks takes the form
\begin{equation}
    \Gq{ij}{y^+}{y}{x^+}{x}{p^+} = \int^{\boldsymbol{r}(y^+) = \boldsymbol{y}}_{\boldsymbol{r}(x^+)=\boldsymbol{x}}\mathcal{D}\boldsymbol{r}(\xi)\Exp{i\frac{p^+}{2}\int_{x^+}^{y^+}ds^+ \, \dot{\boldsymbol{r}}^2}U^{ij}(x^+,y^+,\boldsymbol{r}(\xi)) \, ,
\end{equation}
with the endpoints denoting the starting and final transverse positions of the quark evolution in $x^+$ at fixed light-cone energy $p^+$. We assume $p^+ \gg |\p|$ to be the largest scale in the problem, such that the evolution is highly collimated and all the dynamics are constrained to the plane transverse to the jet axis.\footnote{Of course, particle branching leads to energy sharing between the daughter partons.} The successive scatterings with the medium lead to color precession, which is captured by the path-ordered fundamental Wilson line
\begin{equation}
    U^{ij}(x^+,y^+,\boldsymbol{r}) = \mathcal{P}\Exp{ig\int_{x^+}^{y^+}ds^+\, \cA_a^{-}(s^+, \boldsymbol{r}(s^+))t^a_{ij}}\, .
\end{equation}
Here $\mathcal{A}$ denote the stochastic background field representing the underlying QCD medium. When computing any observable one should average over field configurations, which we assume follow a Gaussian distribution
\begin{equation}\label{pair_correlator} 
    \langle \mathcal{A}_a^-(x^+,x^-,\boldsymbol{x})\mathcal{A}_b^-(y^+,y^-,\boldsymbol{y})\rangle = \delta_{ab}n(x^+)\delta(x^+-y^+)\gamma(\boldsymbol{x-y})\, ,
\end{equation}
with all correlations being local in color, time, and transverse position. Also, $\gamma(\x)$ is directly related to the in-medium elastic scattering rate. Combined with the effective Feynman rules detailed in Appendix~\ref{App1}, one can compute processes such as the one in Fig.~\ref{fig:diagram} resumming all possible gluon exchanges with the medium at leading order in eikonality.

The amplitude describing the top diagram in Fig.~\ref{fig:diagram} reads 
\begin{align}\label{M}
    i\mathcal{M}^{h h'} & = \frac{1}{2q_0^+}e^{i \frac{\boldsymbol{p_1}^2}{2p_1^+}L^+} e^{i \frac{\boldsymbol{p_2}^2}{2p_2^+}L^+}\sum_{\lambda = \pm 1}\int_{x_v^+}\int_{\boldsymbol{x_v, \x_q, \x_{\bar q}, \x_g, \x,\y}} \int_{x_g^+}\mathcal{M}_0^{\lambda, a}(q_0^+, x_g^+,\x_g) e^{-i\boldsymbol{p_1 \cdot x_{q}}} e^{-i\boldsymbol{p_2 \cdot x_{\bar q}}} \nn
    & \times \Gaq{lj}{L^+}{x_{\bar q}}{x_v^+}{x_v}{p_2^+} \left(ig t_{kl}^b V^{\lambda h h'}(z, \boldsymbol{\x}, \boldsymbol{\y})\right)  \Gg{ba}{x_v^+}{x_v-\x}{x_g^+}{x_g}{q_0^+} \nn 
    &\times \Gq{ik}{L^+}{\x_{q}}{x_v^+}{x_v-\y}{p_1^+}\, ,
\end{align}
where
\begin{align}\label{vertex_init_current}
& V^{\lambda h h'}(z, \x, \y) = 2\gamma^{\lambda h}(z)\delta_{h,-h'}\boldsymbol{\epsilon}_{\lambda}\cdot i(\delta(\x)\delta^{'}(\y)+z\delta^{'}(\x)\delta(\y)) \, ,\nn
& \mathcal{M}_0^{\lambda, a}(q_0^+, x_g^+,\x_g) = \int_{p^-,\p}e^{-ip^-x_g^+}e^{i\p\cdot\x_g}\mathcal{M}_0^{\mu,a}(p)\epsilon^{\ast}_{\lambda,\mu}(p) \, ,
\end{align}
and $\mathcal{M}_0^{\mu,a}(p)$ is the momentum space current that produces the gluon. We impose a fixed initial energy for the gluon $p^+ = q_0^+$, and we shall denote the light cone energy fraction carried by the quark by $z\equiv p_1^+/q_0^+$. The notation used for the integrals is, for example:
\begin{equation}
    \int_{x_v^+}\int_{\boldsymbol{x_v}}\int_{\p} = \int_{x_0^+}^{L^+} dx_v^+ \int_{-\infty}^{+\infty}\int_{-\infty}^{+\infty} d^2\boldsymbol{x_v}\int_{-\infty}^{+\infty}\int_{-\infty}^{+\infty} \frac{d^2\p}{(2\pi)^2}  \, .
\end{equation} 
In what follows we shall take $x_0^+=0$.

The vertex $V^{\lambda hh'}$ function in Eq.~\eqref{vertex_init_current} is written in the circular polarization basis and as a function of the transverse polarization of the initial gluon $\lambda = \pm 1$ and of the helicities of the quark ($h = \pm 1$) and anti-quark ($h'=\mp 1$) (cf. Appendix \ref{App1}). The basis for polarization states is, however, not relevant since, as discussed below, we will average over them. As for the fermions' helicity states, these will be discussed further ahead and we keep them explicit since we aim at computing the helicity dependent cross-section. To obtain this usual form for $V^{\lambda hh'}$, we applied the completeness relation $\sum_{\lambda=\pm 1}\epsilon_{\lambda,i}(\p)\epsilon_{\lambda,\mu}^{\ast} = d_{i\mu}(\p)$ to one of the two transverse projectors making up the Lorentz structure of the gluon propagator \cite{Blaizot:2012fh}. To obtain a more familiar expression for the vertex function involving transverse space derivatives, one should recall the property\\
\begin{align}
    \int_{-\infty}^\infty dy \hspace{1mm} \delta'(y)f(x-y) = f'(x)\, ,
\end{align}
which is valid for an arbitrary smooth test function $f$. In the coordinate-space notation used in Eq.~\eqref{M}, the transverse coordinate displacements are associated with the gluon ($\boldsymbol{\x}$) and the quark ($\boldsymbol{\y}$) propagators. The derivatives of $\delta$-functions appearing in $V^{\lambda hh'}$ are transverse vectors which are contracted with the gluon's polarization vector $\boldsymbol{\epsilon}_{\lambda} = \frac{1}{\sqrt{2}}(1,\lambda i)$.

In order to factorize the initial current in light-cone time we assume
\begin{equation}
  \int_{p^-}e^{-ip^-x_g^+}  \mathcal{M}_0^{\mu,a}(p) = \mathcal{M}_0^{\mu,a}(q_0^+,x_g^+,\p)\delta(x_g^+-x_0^+) \, ,
\end{equation}
which means the gluon is produced at the initial time of the medium $x_0^+$ with a fixed energy $q_0^+$. We will stop explicitly writing the initial current's dependence on $q_0^+$ and $x_0^+$ from now on. Also, when squaring the amplitude we take a gluon with a randomised polarization and color (which is fixed to be the same in the conjugate amplitude)
\begin{equation}
    \begin{split}
    & \sum_{\lambda,\bar\lambda = \pm 1}\mathcal{M}_0^{\mu,a}(\p)\epsilon_{\lambda,\mu}^{\ast}\left (\mathcal{M}_0^{\bar\mu,\bar a}(\bar\p)\epsilon_{\bar\lambda,\bar\mu}^{ \ast}\right)^{\dagger} \rightarrow  \sum_{\lambda,\bar\lambda = \pm 1}\frac{\delta^{a\bar a}}{N_c^2-1}\frac{\delta^{\lambda \bar\lambda}}{2}\mathcal{M}_0(\p)\mathcal{M}_0^{\ast}(\bar\p)\, ,
    \end{split}
\end{equation}
where the correction factors below each  Kronecker-$\delta$ account for the average over the fixed choice of quantum numbers. Hence, under these assumptions that impose the production at fixed time of an initial gluon with fixed light-cone energy and quantum numbers, we can work with the following amplitude
\begin{align}\label{eq:M_final}
    i\mathcal{M}^{\lambda h} & = \frac{1}{2q_0^+}e^{i \frac{\boldsymbol{p_1}^2}{2p_1^+}L^+} e^{i \frac{\boldsymbol{p_2}^2}{2p_2^+}L^+}\int_{\p} \mathcal{M}_0(\p)  \int_{x_v^+}\int_{\boldsymbol{x_v, \x_q, \x_{\bar q}, \x_g, \x,\y}} e^{i\boldsymbol{p \cdot x_{g}}}e^{-i\boldsymbol{p_1 \cdot x_{q}}} e^{-i\boldsymbol{p_2 \cdot x_{\bar q}}} \nn
    & \times \Gaq{lj}{L^+}{x_{\bar q}}{x_v^+}{x_v}{p_2^+} \left(ig t_{kl}^b V^{\lambda h}(z, \boldsymbol{\x}, \boldsymbol{\y})\right)  \Gg{ba}{x_v^+}{x_v-\x}{x_0^+}{x_g}{q_0^+} \nn 
    &\times \Gq{ik}{L^+}{\x_{q}}{x_v^+}{x_v-\y}{p_1^+}\, ,
\end{align}
as long as we average over the initial gluon's quantum numbers when squaring. Since the vacuum QCD vertex imposes $h'=-h$, we dropped the explicit dependence on the anti-quark's helicity $h'$.

The color and medium averaged double differential cross-section can be written as
\begin{equation}\label{T}
    4z(1-z)(2\pi)^5\sigma_0\frac{dN^{\lambda h}}{dz d^2\boldsymbol{p_1}d^2\boldsymbol{p_2}} = 2\Re \left[\frac{1}{N_c^2-1}\sum_{\rm colors}\langle  \mathcal{M} \mathcal{M}^{\dagger}\rangle  \right]_{\overline{x}_v^+ > x_v^+} \, ,
\end{equation}
where $\sigma_0$ denotes the initial cross-section to produce the gluon:
\begin{equation}\label{init_xsec}
    \sigma_0 = \int_{\p}\left|\mathcal{M}_0(\p)\right|^2 \, .
\end{equation}
We note that with this choice of normalization, the vacuum limit for this process reads\footnote{Note here the extra $1/2$ factor compared to the usual normalization for the vacuum piece; this can be reabsorbed in the definition of $\sigma_0$.}
\begin{align}\label{true_vac_limit}
    \frac{dN^{h}}{dzd\theta_{q\bar q}d\phi} = \frac{\alpha_s}{(2\pi)^2}\frac{P_{qg}(z)}{\theta_{q\bar q}} \, .
\end{align}

The local form of the medium average in the $+$ component in Eq.~\eqref{pair_correlator} greatly simplifies the structure of the calculation, since it allows to split the overall average into the product of shorter time averages. Using the composition rule for the in-medium propagators
\begin{equation}
    \Gq{ik}{L^+}{x_q}{x_v^+}{x_v-y}{p_1^+} =  \int_{\boldsymbol{z_3}} \Gq{ii_1}{L^+}{x_q}{\overline{x}_v^+}{z_3}{p_1^+}\Gq{i_1k}{\overline{x}_v^+}{z_3}{x_v^+}{x_v-y}{p_1^+}\, ,
\end{equation}
allowing to directly split the medium averages in Eq.~\eqref{T} into three regions: $(x_0^+, x_v^+)$, $(x_v^+, \overline{x}_v^+)$ and $(\overline{x}_v^+, L^+)$. This results in the expression for the cross-section:
\begin{equation}\label{T_expanded}
\begin{split}
    & 4z(1-z)(2\pi)^5\frac{dN^{\lambda h}}{dz d^2\boldsymbol{p_1}d^2\boldsymbol{p_2}}  = \frac{2g^2}{(2q_0^+)^2} \Re \int_{\p, \bar \p} \mathcal{M}_0(\p)\mathcal{M}_0(\bar \p)^{\ast}\\
    & \times \int_{X_v^+}  \int_{\boldsymbol{X}} e^{i\boldsymbol{p \cdot x_g}}e^{-i\boldsymbol{\bar p \cdot \overline{x}_g}} e^{-i\boldsymbol{p_1 \cdot (x_q-\overline{x}_q)}} e^{-i\boldsymbol{p_2 \cdot (x_{\bar q}-\overline{x}_{\bar q})}} V^{\lambda h}(z, \boldsymbol{x}, \boldsymbol{y   }) \left(V^{\lambda h}(z, \boldsymbol{\overline{x}}, \boldsymbol{\overline{y}})\right)^{\ast}\\[2ex]
    & \times \frac{1}{N_c^2-1} \left\langle \Gg{ba}{x_v^+}{x_v-x}{x_0^+}{x_g}{q_0^+} \Ggd{a\overline{b}_1}{x_v^+}{z_1}{x_0^+}{\overline{x}_g}{q_0^+}\right\rangle\\
    & \times t_{kl}^b t_{\overline{l}\overline{k}}^{\overline{b}}  \left\langle \Ggd{\overline{b}_1\overline{b}}{\overline{x}_v^+}{\overline{x}_v-\overline{x}}{x_v^+}{z_1}{q_0^+} \Gaq{lj_1}{\overline{x}_v^+}{z_2}{x_v^+}{x_v}{p_2^+}  \Gq{i_1k}{\overline{x}_v^+}{z_3}{x_v^+}{x_v-y}{p_1+}\right\rangle\\ 
    & \times   \left\langle \Gq{ii_1}{L^+}{x_q}{\overline{x}_v^+}{z_3}{p_1^+} \Gqd{\overline{k}i}{L^+}{\overline{x}_{q}}{\overline{x}_v^+}{\overline{x}_v-\overline{y}}{p_1^+}  \right.\\
    & \times \left. \Gaq{j_1j}{L^+}{x_{\bar q}}{\overline{x}_v^+}{z_2}{p_2^+} \Gaqd{j\overline{l}}{L^+}{\overline{x}_{\bar q}}{\overline{x}_v^+}{\overline{x}_v}{p_2^+} \right\rangle \, , \\
\end{split}
\end{equation}
where, for conciseness, we write $\boldsymbol{X} = \{\boldsymbol{x_v, \overline{x}_v, x_q, \overline{x}_q, x_{\bar q}, \overline{x}_{\bar q}, x_g, \overline{x}_g, x, \overline{x}, y, \overline{y}, z_1, z_2, z_3}\}$ and $X_v^+ = \{x_v^+, \overline{x}_v^+>x_v^+\}$. 

To further simplify Eq.~\eqref{T_expanded}, we perform the remaining two, three and four body averages. To that end, one should project the above object in both the singlet and octet color subspaces. However, since we do not consider the presence of coherent color fields in the medium, the octet contribution should be suppressed and we leave its study to future work, see e.g.~\cite{Barata:2023uoi,Blaizot:2017ypk,Blaizot:1996az,Braaten:1990it} for related discussions. With this in mind, in each time region the system must be in a color singlet state, since Eq.~\eqref{pair_correlator} is local in color space. As a consequence of this statement, we can factorize the color structure in each time region and contract them as detailed in Appendix~\ref{app:2:color_struct}. Using this, the four last lines in Eq.~\eqref{T_expanded} can be written as a product $\mathcal{S}^{(2)}\mathcal{S}^{(3)}\mathcal{S}^{(4)}$, corresponding to the same two, three and four body averages but now projected onto a color singlet state and fully contracted in color. The object $\mathcal{S}^{(2)}$ describes the gluon broadening in the first region and $\mathcal{S}^{(3)}$ the production of the $q\bar q$ pair antenna. The color structure in $\mathcal{S}^{(4)}$ is further simplified by applying the large-$N_c$ limit, in which case it describes the independent broadening of the quark and anti-quark. In Appendix~\ref{regionIApp} and \ref{regionIIIApp} we further simplify these objects by analytically solving the resulting path integrals either fully ($\mathcal{S}^{(2)}$ and $\mathcal{S}^{(4)}$) or partially ($\mathcal{S}^{(3)}$). After carrying out these calculations, an emission \textit{kernel} naturally appears (cf. Appendix~\ref{regionIIIApp})
\begin{equation}
\begin{split}
    \cK(\boldsymbol{u}_1^i,& \boldsymbol{u}_1^f, \boldsymbol{u}_2^i,\boldsymbol{u}_2^f) \equiv \int\mathcal{D}\boldsymbol{u}_1\Exp{\frac{i\tilde{q}_0^+}{2}\int ds^+\dot{\boldsymbol{u}}_1^2}\\
    & \times \Exp{-\frac{1}{2}\int ds^+ \, n(s^+)(N_c (\sigma(\boldsymbol{u}_2^0+z\boldsymbol{u}_1)+\sigma(\boldsymbol{u}_2^0-(1-z)\boldsymbol{u}_1))-\frac{1}{N_c}\sigma(-\boldsymbol{u}_1))}\\
    & \equiv \int_{\boldsymbol{t}_1,\boldsymbol{t}_2,\boldsymbol{t}_3,\boldsymbol{t}_4} e^{i\boldsymbol{t}_1\cdot\boldsymbol{u}_1^i} e^{i\boldsymbol{t}_2\cdot\boldsymbol{u}_1^f} e^{i\boldsymbol{t}_3\cdot\boldsymbol{u}_2^i} e^{i\boldsymbol{t}_4\cdot\boldsymbol{u}_2^f} \tilde{\cK}(\boldsymbol{t}_1,\boldsymbol{t}_2,\boldsymbol{t}_3,\boldsymbol{t}_4) \, ,
\end{split}
\end{equation}
where $\tilde{q}_0^+=z(1-z)q_0^+$ and the full cross-section then reads
\begin{equation}\label{eq:dN_1}
    \begin{split}
        4z(1-z)(2\pi)^5 & \frac{dN^{\lambda h}}{dz d^2\boldsymbol{p_1}d^2\boldsymbol{p_2}} = \frac{g^2}{(2q_0^+)^2} \Re  \int_{\p}|\mathcal{M}_0(\p)|^2
        \int_{X_v^+}\int_{\boldsymbol{v_1, v_2, v_3}}\\
        & \times e^{i\p\cdot\boldsymbol{v_1}} \Exp{-\frac{C_A}{2}\int_{x_0^+}^{x_v^+} ds^+ \, n(s^+)\sigma(\boldsymbol{v_1})}\\
        & \times e^{-i\boldsymbol{p_1 \cdot v_3}} e^{-i\boldsymbol{p_2 \cdot v_2}} \Exp{-\frac{C_F}{2}\int_{\bar x_v^+ }^{L^+} ds^+ \, n(s^+)(\sigma(\boldsymbol{v_3})+\sigma(\boldsymbol{-v_2}))} \\
        & \times \int_{\boldsymbol{t}_1,\boldsymbol{t}_2,\boldsymbol{t}_3,\boldsymbol{t}_4} \tilde{\cK}(\boldsymbol{t}_1,\boldsymbol{t}_2,\boldsymbol{t}_3,\boldsymbol{t}_4)\int_{\boldsymbol{y, \overline{y}, x, \overline{x}, \Delta_v}} V^{\lambda h}(z, \boldsymbol{x}, \boldsymbol{y}) \left(V^{\lambda h}(z, \boldsymbol{\overline{x}}, \boldsymbol{\overline{y}})\right)^{\ast}\\
        & \times \left(\frac{q_0^+}{2\pi\Delta t}\right)^2 \Exp{\frac{iq_0^+}{2}\frac{\Delta\boldsymbol{u}_2}{\Delta t}\left(2\Delta\boldsymbol{r}_0 + \Delta\boldsymbol{u}_2\right)} 
       e^{i\boldsymbol{t}_1\cdot\boldsymbol{u}_1^i} e^{i\boldsymbol{t}_2\cdot\boldsymbol{u}_1^f} e^{i\boldsymbol{t}_3\cdot\boldsymbol{u}_2^i} e^{i\boldsymbol{t}_4\cdot\boldsymbol{u}_2^f}  \, .
    \end{split}
\end{equation}
Here we used the notation $\Delta \boldsymbol{X} = \boldsymbol{X^f} - \boldsymbol{X^i}$ and, for completeness, we have $\Delta\r_0 = \x+\boldsymbol{v_1}-\bar\x+\boldsymbol{\Delta_v}$ and
\begin{equation}
    \begin{split}
        & \boldsymbol{u_1^i} = \boldsymbol{y}\\
        & \boldsymbol{u_2^i} = \boldsymbol{v_1}+\x-z\boldsymbol{y}
    \end{split}
    \qquad \qquad
    \begin{split}
        & \boldsymbol{u_1^f} = \boldsymbol{v_2}-\boldsymbol{v_3}+\boldsymbol{\overline{y}} \\
        & \boldsymbol{u_2^f} = \boldsymbol{v_2}+\overline{\x}-z(\boldsymbol{v_2}-\boldsymbol{v_3}+\boldsymbol{\overline{y}}) \, .
    \end{split}
\end{equation}
 Note that we have changed variables from $\bar\x_v$ to $\boldsymbol{\Delta_v} = \bar\x_v - \x_v$ and integrated over $\x_v$ and $\bar \p$, as explained at the end of Appendix~\ref{regionIIIApp}. This sets $\p = \bar\p$, i.e., the same transverse momentum flows out of the initial current in amplitude and its complex conjugate.

Let us take note on an important aspect regarding phase space dependence. It is an intuitive idea that one should keep the spectrum differential in some directional quantity in order to be able to probe anisotropy induced modifications. However, as we will see, integrating over all momenta still leaves trace of the medium anisotropy. Despite this, we choose to keep the spectrum differential in some angle(s) which we will define ahead, since this provides an additional handle on probing anisotropic effects. In fact, keeping some differential dependence on an angle in the transverse plane is necessary to keep spin/helicity dependence. To that end, we switch our phase space variables from $\{\boldsymbol{p_1}, \boldsymbol{p_2}\}$ to $\{\boldsymbol{P}^{CM}, \boldsymbol{P}^{rel}\}$, where the center of transverse momentum and the relative transverse momentum are, respectively, given by

\begin{align}
        & \boldsymbol{P}^{CM} = \boldsymbol{p_1} + \boldsymbol{p_2} \, ,\quad  \boldsymbol{P}^{rel} = (1-z)\boldsymbol{p_1}-z\boldsymbol{p_2} \, ,
\end{align}
which is a transformation with unit jacobian. $\boldsymbol{P}^{rel}$ ($-\boldsymbol{P}^{rel}$) is the momentum of the quark (anti-quark) in the center of momentum frame. 

 We now carry out the derivatives in the vertex, integrate out the center of momentum information $\boldsymbol{P}^{CM}$ and at the same time average over gluon polarizations  $\lambda$. By using the massless, helicity-dependent vertex in Eq.~\eqref{vertex_init_current}, this leads us to: 
\begin{equation}\label{eq:help1}
    \begin{split}
        & 4z(1-z)(2\pi)^3\frac{dN^{h}}{dz d^2\boldsymbol{P}^{rel}} = -
        \frac{2g^2}{(2q_0^+)^2z(1-z)}\Re \int_{X_v^+}\int_{\boldsymbol{v}}\int_{\boldsymbol{t}_1,\boldsymbol{t}_2}\\
        & \times e^{-i(\boldsymbol{P}^{rel}+\boldsymbol{t}_2) \cdot \boldsymbol{v}} \Exp{-\frac{C_F}{2}\int_{\bar x_v^+}^{L^+} ds^+ \, n(s^+)(\sigma((1-z)\boldsymbol{v})+\sigma(z\boldsymbol{v}))}\\
        & \times \left[P_{qg}^{vac}(z)\left(\boldsymbol{t}_1\cdot \boldsymbol{t}_2\right)+\frac{ih}{2}(1-2z)\left(\boldsymbol{t}_1\times\boldsymbol{t}_2\right)_z\right]  \tilde{\cK}(\boldsymbol{t}_1,\boldsymbol{t}_2) \, ,
    \end{split}
\end{equation}
where 
\begin{equation}
\begin{split}
    & \tilde{\cK}(\boldsymbol{t}_1, \boldsymbol{t}_2) = \int_{\boldsymbol{a_1}, \boldsymbol{a_2}} e^{-i\boldsymbol{a_1}\cdot\boldsymbol{t}_1} e^{-i\boldsymbol{a_2}\cdot\boldsymbol{t}_2}\int_{\boldsymbol{a_1}}^{\boldsymbol{a_2}}\mathcal{D}\boldsymbol{u}\Exp{\frac{i\tilde{q}_0^+}{2}\int_{x_v^+}^{\bar x_v^+} ds^+\dot{\boldsymbol{u}}^2}\\
    & \times \Exp{-\frac{1}{2}\int_{x_v^+}^{\bar x_v^+} ds^+ \, n(s^+)(N_c (\sigma(z\boldsymbol{u})+\sigma(-(1-z)\boldsymbol{u}))-\frac{1}{N_c}\sigma(-\boldsymbol{u}))} \, ,
\end{split}
\end{equation} 
and $\left(\boldsymbol{t}_1\times\boldsymbol{t}_2\right)_z = \boldsymbol{t}_{1x}\boldsymbol{t}_{2y}-\boldsymbol{t}_{1y}\boldsymbol{t}_{2x}$. Note that this integration in $\boldsymbol{P}^{CM}$ allowed us to factorize the integral over $\p$ of the initial current squared ($\sigma_0$ in Eq.~\eqref{init_xsec}) from the remaining of the amplitude, so we have dropped it. This integration also resulted in the gluon broadening being completely integrated out.

At this point we have to provide an explicit form for the dipole cross-section $\sigma$. Here we use the harmonic approximation~\cite{Zakharov:1996fv}, whereby the logarithmic dependence on the transverse position, in the small distance limit, is dropped in $\sigma$, see e.g.~\cite{Mehtar-Tani:2019tvy,Barata:2020sav} for further discussion on the limitations of this approximation. Additionally assuming that the medium is static, we can write 
\begin{equation}
        N_c \int_{x^+}^{y^+} ds^+ \, n(s^+) \sigma(\boldsymbol{r}) = N_c  (y^+-x^+) n \sigma(\boldsymbol{r}) = \frac{(y^+-x^+)}{2}\left(\hat{q}_x r_x^2 + \hat{q}_y r_y^2\right) \, ,
\end{equation}
where we have written the components of the transverse vector as $\r = (r_x,r_y)$.  Furthermore, we have naturally introduced a form of medium anisotropy by considering different values for the jet quenching parameter $\hat q$ in different directions in the transverse plane \cite{Hauksson:2023tze}. Combining this with Eq.~\eqref{eq:help1}, we obtain
\begin{equation}\label{before_derivatives_massless}
\begin{split}
    & 4z(1-z)(2\pi)^3\frac{dN^{h}}{dz d^2\boldsymbol{P}^{rel}} = 
    \frac{2g^2}{(2q_0^+)^2z(1-z)}\\
    & \times \Re  \int_{X_v^+}\int_{\boldsymbol{v}} e^{-i\boldsymbol{P}^{rel} \cdot \boldsymbol{v}} \Exp{-\frac{C_F}{2N_c}\Delta_L^+ P_{qg}^{vac}(z)\left(\hat{q}_x v_{x}^2 + \hat{q}_y v_{y}^2\right)} \\[1ex]
    & \times \left[P_{qg}^{vac}(z)\left(\boldsymbol{\nabla_{a_1}}\cdot \boldsymbol{\nabla_{a_2}}\right)+\frac{ih}{2}(1-2z)\left(\boldsymbol{\nabla_{a_1}}\times\boldsymbol{\nabla_{a_2}}\right)_z\right]  \cK(\boldsymbol{a_1},\boldsymbol{a_2})\Big{|}_{\boldsymbol{a_1}=0, \boldsymbol{a_2}=-\boldsymbol{v}} \, ,
\end{split}
\end{equation} 
where $\Delta_L^+ = L^+ - \overline{x}_v^+$. Importantly, note that if the medium was isotropic there would be no helicity ($h$) dependence attached to medium parameters $\hat{q}_x, \hat{q}_y$. The path integral $\cK(\boldsymbol{a_1}, \boldsymbol{a_2})$ has a well known solution in the literature \cite{Apolinario:2014csa, Kleinert:2004ev}
\begin{equation}\label{HO_pathint}
\begin{split}
     \cK(\boldsymbol{a_1},\boldsymbol{a_2}) &=  \sqrt{\frac{\tilde{q}_0^+\Omega_x}{2\pi i \sin{\Omega_x\Delta t}}}\Exp{i \frac{\tilde{q}_0^+\Omega_x}{2 \sin{\Omega_x\Delta t}} \left[ \left(a_{1x}^2+a_{2x}^2\right)\cos{\Omega_x\Delta t} - 2a_{1x}a_{2x}\right]} \\
    & \times \sqrt{\frac{\tilde{q}_0^+\Omega_y}{2\pi i \sin{\Omega_y\Delta t}}} \Exp{i \frac{\tilde{q}_0^+\Omega_y}{2 \sin{\Omega_y\Delta t}} \left[\left(a_{1y}^2+a_{2y}^2\right)\cos{\Omega_y\Delta t} - 2a_{1y}a_{2y}\right]} \, ,
\end{split}
\end{equation} 
with 
\begin{equation}
    \Omega_i = \left(\frac{1-i}{\sqrt{2}}\right)\omega_i, \qquad \omega_i \approx  \sqrt{\frac{\hat{q}_iP_{qg}^{vac}(z)}{\tilde{q}_0^+}},  \qquad \Delta t = \overline{x}_v^+ - x_v^+ \, ,
\end{equation}
where $P_{qg}^{vac}(z)= T_R (z^2+ (1-z)^2)$ is the vacuum splitting function, we took $C_F/N_c \approx 1/2$ and we ignored a $1/N_c^2$ term in $\omega_i$.

We now choose to write the result in terms of two angles: the angle between the 3-momenta $\vec{p}_1$ and $\vec{p_2}$ and the azimuthal angle of the relative transverse momentum $\boldsymbol{P}^{rel}$. These are given by\\
\begin{align}
    & |\boldsymbol{P}^{rel}|= \sin\left(\frac{\theta_{q\bar q}}{2}\right) \tilde{q}_0^+\sqrt{2} \approx \frac{\theta_{q\bar q}}{\sqrt{2}} \tilde{q}_0^+ \, ,\quad      \phi = \arctan \frac{\P^{rel}_y}{\P^{rel}_x} \, ,
\end{align}
where the expression for $\theta_{q\overline{q}}$ assumes $(|\p_i|/p_i^+)^2 \ll 1$. Using the results in Eq.~\eqref{before_derivatives_massless} and \eqref{HO_pathint}, one can then write the differential cross-section as a function of these two angles as
\begin{align}\label{azimuth_opening_angle_final}
     &\frac{dN^{h}}{dzd\theta_{q\bar q}d\phi}=    \frac{\alpha_s \theta_{q\bar q}}{8\pi^2}  \text{Im}  \int_{X_v^+} \frac{\sqrt{c_{1x}}\sqrt{c_{1y}}}{\sqrt{c_{3x}}\sqrt{c_{3y}}} \Exp{i\frac{\tilde{q}_0^+\theta^2_{q\bar q}}{4}\left( \frac{\cos^2\phi}{c_{3x} }+ \frac{\sin^2\phi}{c_{3y}}\right)}\nn
    & \times \Bigg[P_{qg}^{vac}(z)\Bigg(\tilde{q}_0^+c_4 +\tilde{q}_0^+\left(\frac{c_{1x}c_{2x}}{c_{3x}}+\frac{c_{1y}c_{2y}}{c_{3y}}\right) + i\frac{(\tilde{q}_0^+)^2\theta^2_{q\bar q}}{2}\left(\frac{c_{1x}c_{2x}}{c_{3x}^2}\cos^2\phi + \frac{c_{1y}c_{2y}}{c_{3y}^2}\sin^2\phi \right) \Bigg)\nn
    & + \frac{h(1-2z)}{4}(\tilde{q}_0^+)^2\theta^2_{q\bar q}\frac{c_5 \cos\phi\sin\phi}{c_{3x}c_{3y}}\Bigg] \, ,
\end{align}
 where we have introduced the compact notation
\begin{equation}\label{ci_definition}
    \begin{split}
    &c_{1i} = \frac{\Omega_i}{2i\sin{\Omega_i\Delta t}}, \qquad \qquad c_{2i} = \frac{\Omega_i}{\tan{\Omega_i\Delta t}}, \qquad c_{3i} = \frac{\Delta_L^+\Omega_i^2}{2}-c_{2i}, \\
    &c_4 = \left(c_{1x}+c_{1y}\right), \hspace{1 cm} \qquad  c_5 = \left(c_{1y}c_{2x}-c_{1x}c_{2y}\right) \, .
    \end{split}
\end{equation}
We should emphasize that if we were to integrate over $\boldsymbol{P}^{rel}$ in Eq.~\eqref{before_derivatives_massless} the only term left would be the one proportional to $c_4$, which is nevertheless sensitive to an anisotropic $\hat q$. That term would be similar to the one obtained for the decay rate of a $y$-polarized gluon to two $z-$polarized gluons in \cite{Hauksson:2023tze}, in which case a (linear) polarization dependence exists even after integrating over all momenta. Finally, the purely azimuthal distribution is given by\footnote{It is worth pointing out that $\text{Im}(c_{3i}) < 0$ for all $x_v^+$ and $\bar x_v^+$ guarantees the convergence of both integrals in $\boldsymbol{v}$ and $|\boldsymbol{P}^{rel}|$ and is responsible for the sign attached to the second term in Eq.~\eqref{azimuth_opening_angle_final}.}
\begin{equation}\label{azimuth_final}
    \begin{split}
    & \frac{dN^{h}}{dzd\phi} = 
    \frac{\alpha_s}{(2\pi)^2} \Re  \int_{X_v^+} \frac{\sqrt{c_{1x}}\sqrt{c_{1y}}}{\sqrt{c_{3x}}\sqrt{c_{3y}}} \left(P_{qg}^{vac}(z)\Bigg(\frac{(c_{1y}c_{2y}c_{3x}-c_{1x}c_{2x}c_{3y})(c_{3y}\cos^2{\phi}-c_{3x}\sin^2{\phi})}{\left(c_{3y}\cos^2{\phi}+c_{3x}\sin^2{\phi}\right)^2}\right. +\\
    &  \left.+\frac{c_{3x}c_{3y}c_4}{c_{3y}\cos^2{\phi}+c_{3x}\sin^2{\phi}}\Bigg) + ih(1-2z) \frac{ c_{3x}c_{3y}c_5\sin{\phi}\cos{\phi}}{\left(c_{3y}\cos^2{\phi}+c_{3x}\sin^2{\phi}\right)^2}\right) \, .
    \end{split}
\end{equation}
Note that the distribution in Eq.~\eqref{azimuth_final} has a linear divergence $1/\Delta t$ as $\Delta t \rightarrow 0$, as noted in~\cite{Hauksson:2023tze}, which is removed by subtracting a contribution corresponding to $\hat q_i \rightarrow 0$ to all our results.

A similar calculation for the case of massive final states can also be performed, which we here largely omit and refer the reader to Appendix~\ref{App1}. The most relevant modification when introducing masses, while still working in the high energy limit, is the appearance of a mass dependent term where the spins of the quark ($r$) and anti-quark ($s$) can be anti-parallel in the vertex (see Eq.~\eqref{massive_squared}). An additional phase factor proportional to $m^2$ (cf. Eq.~\eqref{massive_phase}) is also generated by the mass term. The polarization averaged result is given by:
\begin{equation}\label{massive_zaligned_final}
    \begin{split}
    & \frac{dN^{r s}}{dzd\phi} = \frac{\alpha_s}{(2\pi)^2} \Re  \int_{X_v^+} \frac{\sqrt{c_{1x}}\sqrt{c_{1y}}}{\sqrt{c_{3x}}\sqrt{c_{3y}}} e^{-i\frac{m^2}{2\tilde{q}_0^+}\Delta t}\\
    & \times \Bigg\{\delta_{rs}\left(P_{qg}^{vac}(z)\left[\frac{(c_{1y}c_{2y}c_{3x}-c_{1x}c_{2x}c_{3y})(c_{3y}\cos^2{\phi}-c_{3x}\sin^2{\phi})}{\left(c_{3y}\cos^2{\phi}+c_{3x}\sin^2{\phi}\right)^2}\right.\right. +\\
    & \left.\left.+\frac{c_{3x}c_{3y}c_4}{c_{3y}\cos^2{\phi}+c_{3x}\sin^2{\phi}}\right] + ir(1-2z) \frac{ c_{3x}c_{3y}c_5\sin{\phi}\cos{\phi}}{\left(c_{3y}\cos^2{\phi}+c_{3x}\sin^2{\phi}\right)^2}\right)\\
    & \delta_{r,-s}\left(\frac{m^2}{4\tilde{q}_0^+}\frac{c_{3x}c_{3y}}{c_{3y}\cos^2{\phi}+c_{3x}\sin^2{\phi}}\right)\Bigg\} \, ,
    \end{split}
\end{equation}
where now we use the quark's spin-projection along $z$ rather than the helicity $h$, since we are considering finite masses.

\section{Jet observables}\label{sec:observables}
In the previous section, we provided the cross-section to produce a quark antenna in a dense anisotropic medium, for both massless and massive final states. Using these results, we now study the impact of the medium anisotropy, encapsulated into $\hat q$, on jet observables. We will first study the total particle distributions. This is complemented by a study of the obtained observables in terms of their Fourier decomposition, similar to the extraction of flow harmonics commonly performed in soft physics studies in URHICs, see e.g.~\cite{Snellings:2011sz}. Finally we discuss how measurements of spin along different directions in the transverse plane can give information about jet polarization.

\subsection{Azimuthal particle distribution}
We first consider the azimuthal particle distribution given by Eqs.~\eqref{azimuth_final} and ~\eqref{massive_zaligned_final}. The massless case only has one possible helicity configuration ($\delta_{h',-h}$) as we have seen and for the massive case we will focus on the same spin configuration ($\delta^{rs}$), given by the first term  in Eq.~\eqref{massive_zaligned_final}. The numerical results are presented in terms of the dimensionless quantities
\begin{equation}\label{dimensionless_var_def}
    \zeta = \frac{\sqrt{\hat q_y} - \sqrt{\hat q_x}}{\sqrt{\hat q_y} + \sqrt{\hat q_x}}\, ,\quad r = L^+ \frac{\left(\sqrt{\hat q_y} + \sqrt{\hat q_x}\right)}{2\sqrt{q_0^+}} \, , \qquad \mu = \frac{\sqrt{2} m^2}{\left(\sqrt{\hat q_y} + \sqrt{\hat q_x}\right)\sqrt{q_0^+}} \, ,
\end{equation}
which give a measure of the anisotropy $\zeta$, a rescaled medium length $r$ and a rescaled quark mass $\mu$.
For the numerical evaluations we estimate the values of these parameters by considering 
$L^+/\sqrt{2}= 10$ fm, $q_0^+/\sqrt{2}=100$ GeV, and $\hat q_x=1 \, {\rm GeV}^2 {\rm fm}^{-1}$. Writing $\hat q_y=c \hat q_x$ and using $m=4$ GeV/c$^2$, we have that
\begin{align}
   &c=2: \, \zeta=  0.17 \, , r = 3.20 \, , \mu= 1.76 \, , \nn 
    &c=20 : \, \zeta=  0.63 \, , r = 7.28 \, , \mu= 0.78\, .
\end{align}
Thus, in what follows we take $0.10\leq\zeta\leq 0.95$, $r=5$, and $\mu=0.1,0.4$.\footnote{To better illustrate the anisotropic effects, in the numerical study we used slightly smaller values for $\mu$, since the mass suppresses the medium signature. This can be cleanly seen in the results shown.}

\begin{figure}
     \centering
     \begin{subfigure}[h]{0.49\textwidth}
         \centering
         \includegraphics[width=\textwidth]{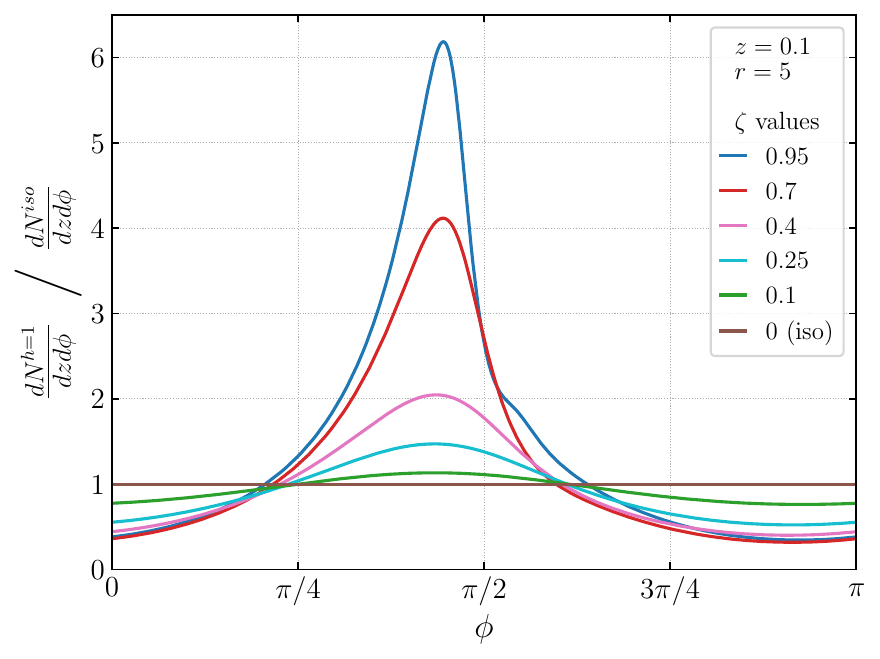}
     \end{subfigure}
     \begin{subfigure}[h]{0.49\textwidth}
         \centering
         \includegraphics[width=\textwidth]{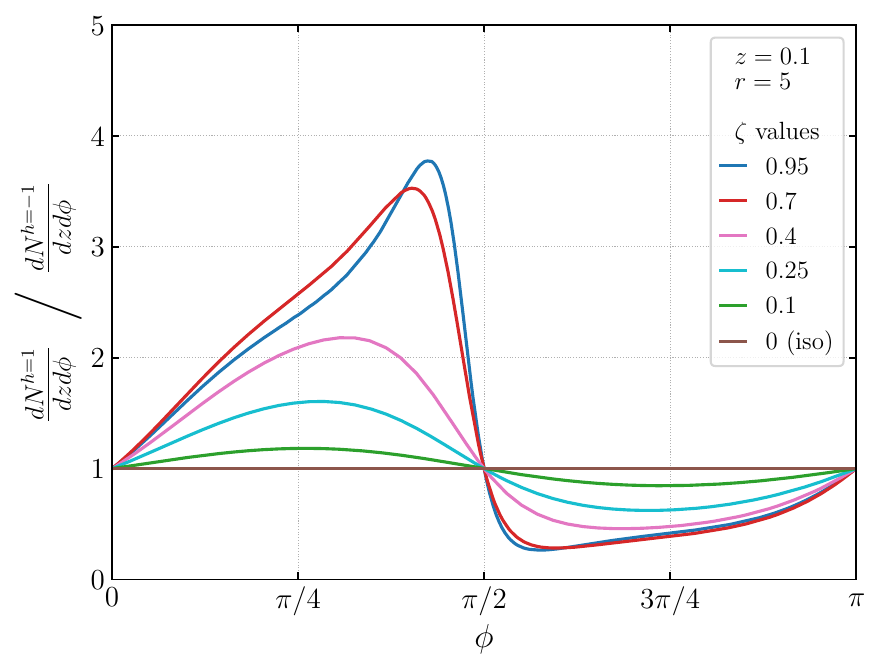}
     \end{subfigure}
     \begin{subfigure}[h]{0.49\textwidth}
         \centering
         \includegraphics[width=\textwidth]{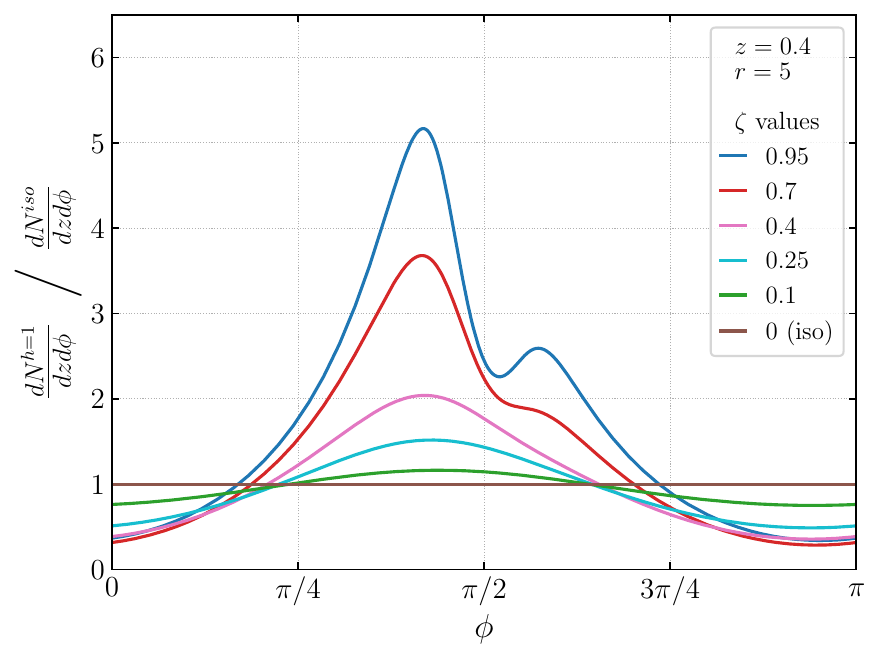}
     \end{subfigure}
     \begin{subfigure}[h]{0.49\textwidth}
         \centering
         \includegraphics[width=\textwidth]{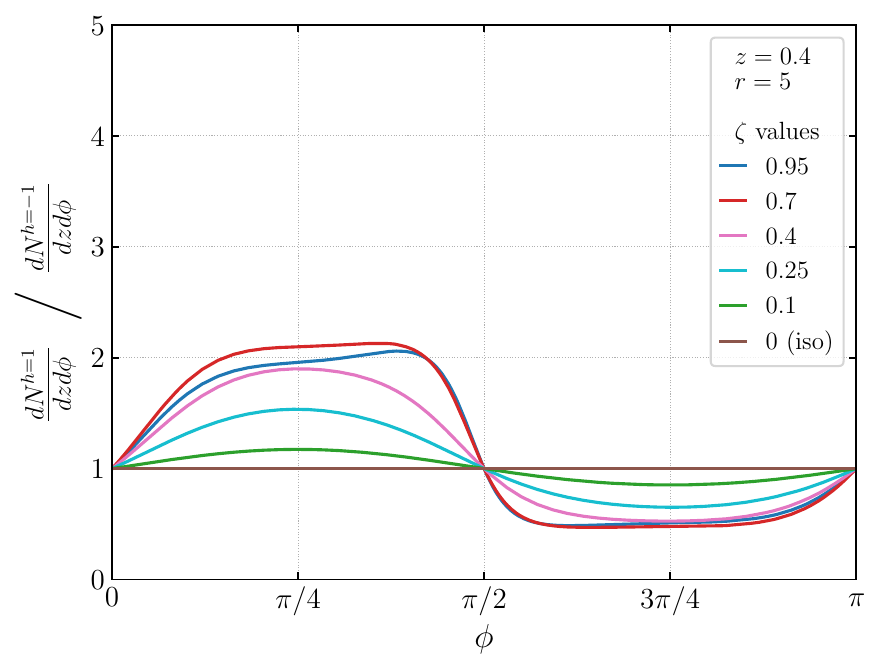}
     \end{subfigure}
     \caption{Particle number distribution (Eq.~\eqref{azimuth_final}) as a function of $\phi$ for different values of the anisotropy parameter $\zeta$ and for fixed $r=5$. \textbf{Left:} ratio of the $h=1$ case with respect to the isotropic ($\zeta = 0$) case. \textbf{Right:} ratio of the $h=1$ case with respect to the $h=-1$ case. The top row corresponds to $z=0.1$ and bottom to $z=0.4$. The vacuum piece is subtracted.}
     \label{azimuthal_plot_phi}
\end{figure}

\begin{figure}
     \centering
     \begin{subfigure}[h]{0.49\textwidth}
         \centering
         \includegraphics[width=\textwidth]{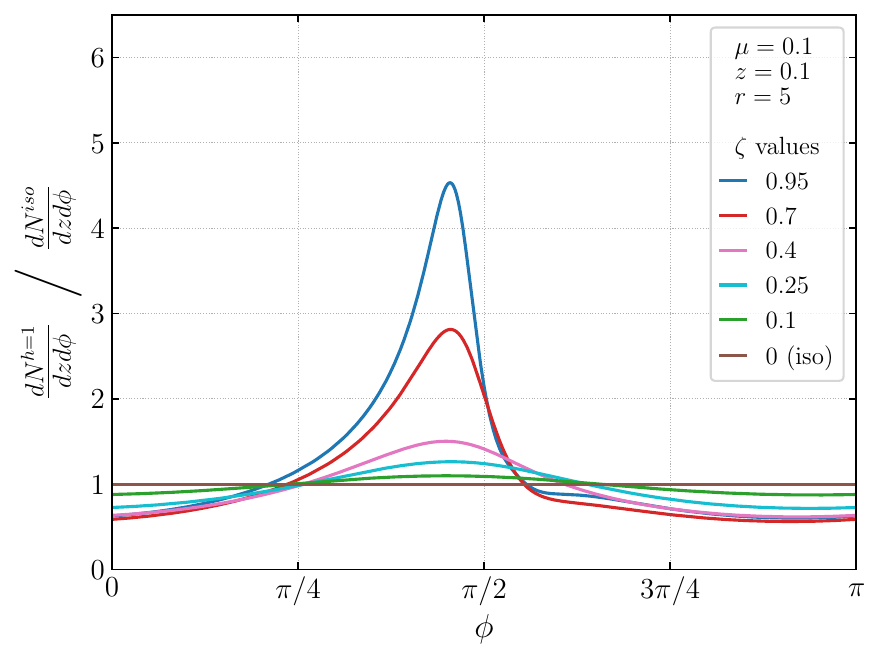}
         \label{fig:three sin x}
     \end{subfigure}
     \begin{subfigure}[h]{0.49\textwidth}
         \centering
         \includegraphics[width=\textwidth]{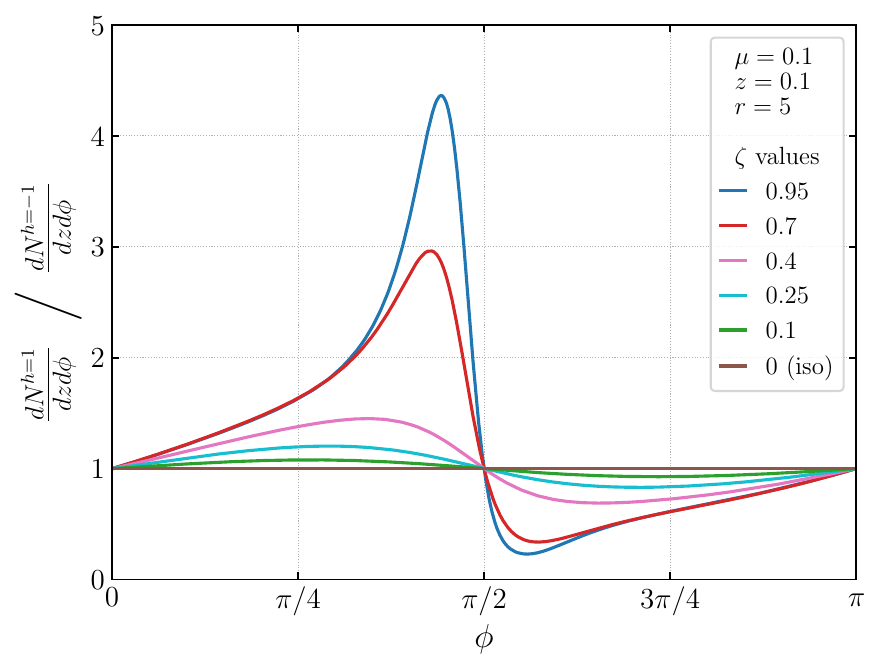}
         \label{fig:three sin x}
     \end{subfigure}
     \begin{subfigure}[h]{0.49\textwidth}
         \centering
         \includegraphics[width=\textwidth]{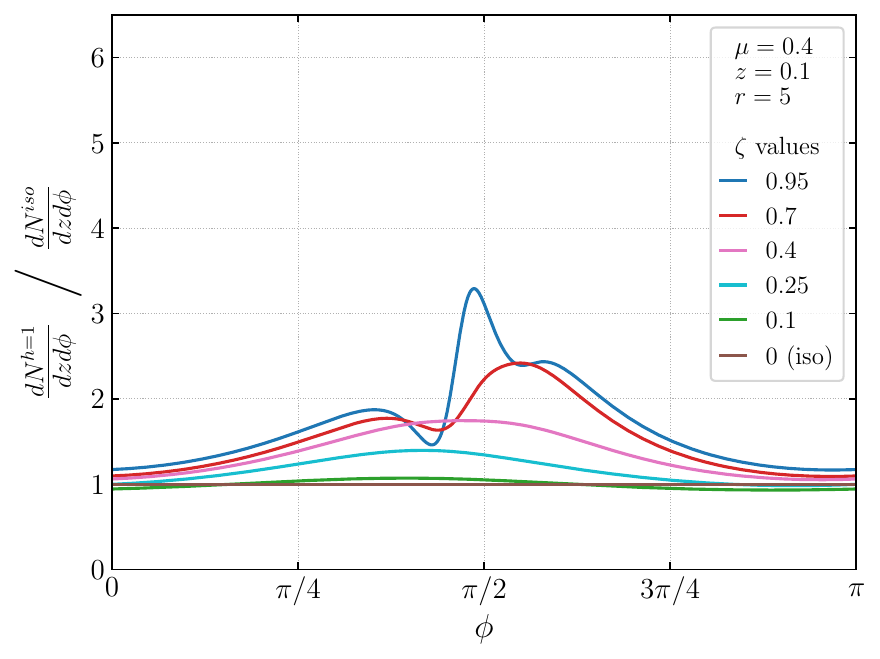}
         \label{fig:three sin x}
     \end{subfigure}
     \begin{subfigure}[h]{0.49\textwidth}
         \centering
         \includegraphics[width=\textwidth]{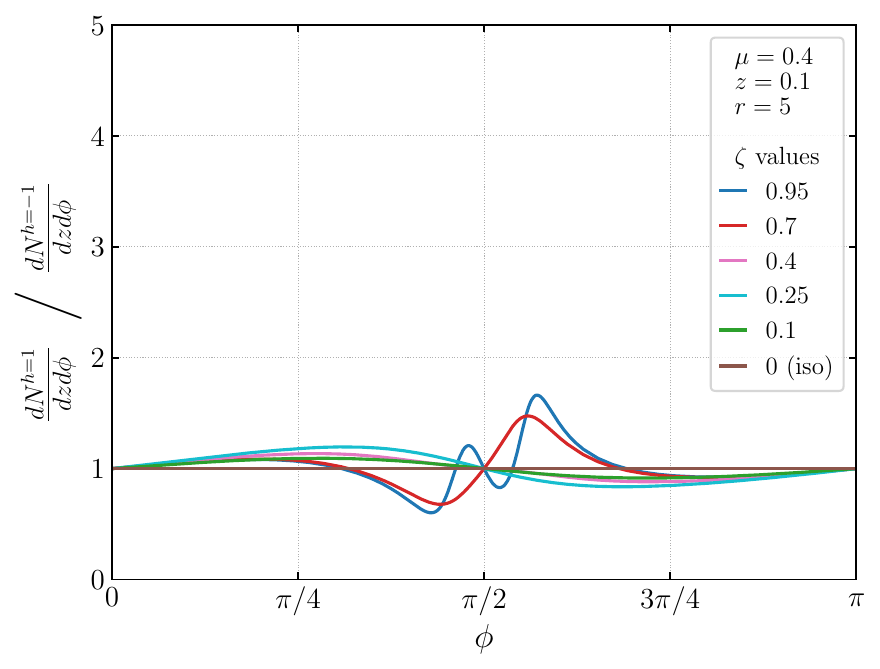}
         \label{fig:three sin x}
     \end{subfigure}
     \caption{Particle number distribution (Eq.~\eqref{massive_zaligned_final}) as a function of $\phi$ for different values of the anisotropy parameter $\zeta$ and for fixed $r=5$ and $z=0.1$. Note that the spin index $r$ is renamed to $h$. \textbf{Left:} ratio of the $h=1$ case with respect to the isotropic ($\zeta = 0$) case. \textbf{Right:} ratio of the $h=1$ case with respect to the $h=-1$ case. The top row corresponds to $\mu=0.1$ and bottom to $\mu=0.4$. The vacuum piece is subtracted.}
     \label{massive_azimuthal_plot_phi}
\end{figure}

In Fig.~\ref{azimuthal_plot_phi} we show the evolution of the particle distribution for massless quark as a function of $\phi$ for several values of the anisotropic coefficient $\zeta$. The result for massive quarks is shown in Fig.~\ref{azimuthal_plot_phi}.
The left columns shows the ratio between the $h=1$ distribution and the isotropic one ($\zeta=0$). The isotropic distribution is flat ($\phi$ independent), so the ratio is merely a change in normalization of all curves. In both massless and massive cases we see that a larger anisotropy results in a larger ratio near $\pi/2$, where the distinction with respect to the isotropic case seems the largest. Additionally, there is a small suppression near $0$ and $\pi$ which only slightly increases with the value of the anisotropy. For $\hat q_x > \hat q_y$, i.e. $\zeta < 0$, we would obtain the reverse behaviour -- peaks around $0$ and $\pi$.
We further observe that the distribution is approximately symmetric with respect to $\phi=\pi/2$, which indicates that the helicity independent terms (symmetric for $\phi\rightarrow \pi-\phi$) dominate over the dependent ones (anti-symmetric for the same transformation). This results in an elliptic-type deformation of the particle distribution in the transverse plane. On the right columns we show the ratio between the $h=\pm 1$ distributions, which quantifies the importance of the helicity dependent term. One can distinguish the two helicity states better the greater the anisotropy and the softest the splitting (owing to the $(1-2z)$ factor) and this is again more pronounced around $\pi/2$. The symmetric behaviour is purely a consequence of two distributions corresponding to $h=\pm 1$ being mirrored with respect to $\phi=\pi/2$. The two peaks being in distinctly different regions indicates that positive and negative helicity states will populate different regions inside the jet. Regarding the massive case in specific, it seems that including mass reduces the sensitivity to anisotropy values (Fig.~\ref{massive_azimuthal_plot_phi} on the left) and that the distinction between the two spin states becomes much less pronounced (Fig.~\ref{massive_azimuthal_plot_phi} on the right).

\subsection{Fourier decomposition}
The different symmetry properties of the two contributions, spin/helicity dependent and independent, to the final state distribution suggests that one can isolate the helicity terms through a harmonic Fourier decomposition. We can further motivate this decomposition by performing an expansion of the distribution in powers of $\zeta$. We find at leading order
\begin{equation}\label{small_anisotropy_expansion}
    \begin{split}
        & \frac{dN^h}{dzd\phi} = \frac{\alpha_s}{2\pi} \Re \int_{X_v^+} e^{-i\frac{m^2}{2\tilde{q}_0^+}\Delta t}\Bigg[2c_1^2P_{qg}^{vac}(z) + \frac{2c_1^2}{c_3}\,\zeta\Bigg(ih\frac{1-2z}{2}\bar\Omega^2\Delta t\sin(2\phi) + \\
        & P_{qg}^{vac}(z)\left(\frac{4c_1^2(c_2+c_3)\Delta t-c_2^2(1+c_3\Delta t)}{c_3} - (c_2+2c_3-4c_1^2\Delta t)  \right)\cos(2\phi)\Bigg)\Bigg]
    \end{split}
\end{equation}
where we used a new shorthand notation
\begin{equation}\label{expand_ci_definition}
    \begin{split}
    &c_{1} = \frac{\bar\Omega}{2i\sin{\bar\Omega\Delta t}} \, , \quad \qquad c_{2} = \frac{\bar\Omega}{\tan{\bar\Omega\Delta t}}\, , \quad c_{3} = \frac{\Delta_L^+\bar\Omega^2}{2}-c_{2}\, ,
    \end{split}
\end{equation}
where $\bar\Omega = \left(\frac{1-i}{\sqrt{2}}\right)\bar \omega$, and $\bar \omega = (\omega_y+\omega_x)/2$. Equation~\eqref{small_anisotropy_expansion} neatly illustrates that, at leading power in $\zeta$, the helicity independent terms are proportional to $\cos(2\phi)$, i.e. the elliptic contribution, while the helicity dependence goes with $\sin(2\phi)$, as argued above. In fact, this observation can be promoted to all orders in $\zeta$. Performing the Fourier decomposition for finite $\zeta$ then yields
\begin{align}\label{harmonic_exp}
   \frac{2\pi}{ dN^h/dz} \frac{dN^{h}}{dzd\phi}  = 1+ \sum_{n=1}^\infty v^{(h)}_n \cos\left(n \phi \right) + \sum_{n=1}^\infty w^{(h)}_n \sin(n\phi) \, , 
\end{align}
where $v_n\equiv (dv_n^{(h)}/dz)/(dv_0^{(h)}/dz) $, $w_n = (dw_{n}^{(h)}/dz)/(dv_0^{(h)}/dz)$, and we have the relations
\begin{align}
v_0^{(h)} &= \frac{1}{2\pi} \int_0^{2\pi} d\phi \,\,  2\pi \frac{dN^{h}}{dz d\phi} \, , \nn 
v_n^{(h)} &= \frac{1}{\pi} \int_0^{2\pi} d\phi \,\,  2\pi \frac{dN^{h}}{dzd\phi} \, \cos\left(n \phi \right) \, ,\nn 
w_n^{(h)}&= \frac{1}{\pi} \int_0^{2\pi} d\phi \, \,    2\pi\frac{dN^{h}}{dzd\phi} \, \sin\left( n\phi \right)\, .
\end{align}  
At this point, it is important to note that several of these Fourier coefficients can be immediately determined from the symmetries of the spectrum in Eq.~\eqref{azimuth_final}. First of all, in the case of an isotropic medium, all Fourier coefficients except for $v_0$ vanish. 
For an anisotropic medium, both $v^{(h)}_n$ and $w_n^{(h)}$ vanish for odd $n$ due to discrete symmetry $\phi \rightarrow \phi - \pi$ of the particle distribution. This is expected since the presence of the anisotropic $\hat q$ can not lead to any direct flow-like contribution. The coefficients $v_n$ for even $n$ are non-vanishing but independent of helicity, i.e., $v^{(h)}_n = v_n$, a consequence of $\phi \rightarrow -\phi$ antisymmetry of the helicity dependent term. The helicity dependence can only enter via the odd-parity terms $w_n^{(h)}$ in the series. This is analogous to charge imbalance measurements in chiral magnetic effect studies, see e.g.~\cite{Kharzeev:2015znc}.\footnote{Although we note that in our context we do not need to look into fluctuations, as is the case in studies of the chiral magnetic effect in heavy ions.} It is also easy to realize that  $w_n^{(+1)}=-w_n^{(-1)}$, since $w_n^{(h)} \propto h$. Thus, the presence of non-zero values of any of these coefficients gives information about the underlying properties of the medium and its coupling to the jet in different ways.

\begin{figure}
     \centering
     \begin{subfigure}[h]{0.49\textwidth}
         \centering
         \includegraphics[width=\textwidth]{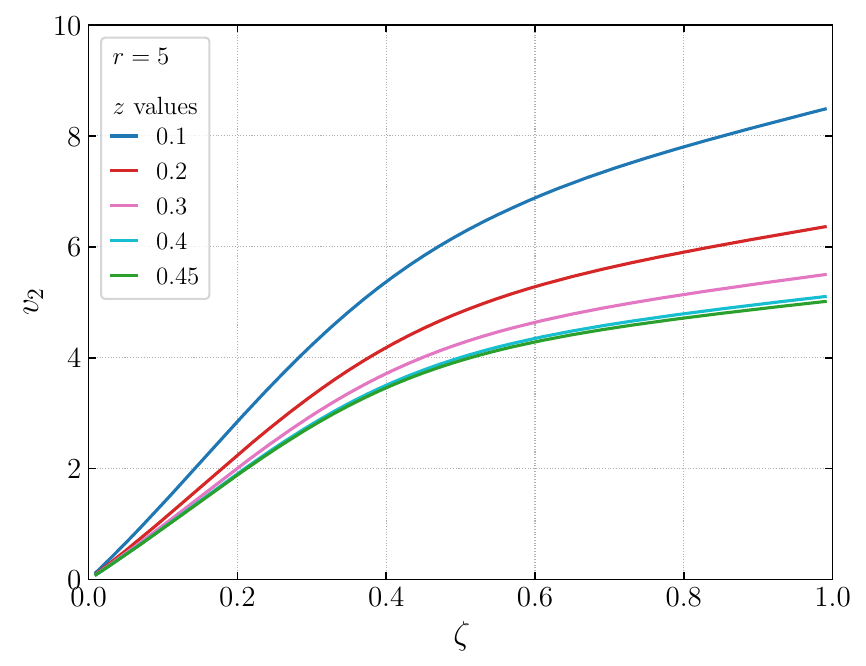}
         \label{fig:three sin x}
     \end{subfigure}
     \begin{subfigure}[h]{0.49\textwidth}
         \centering
         \includegraphics[width=\textwidth]{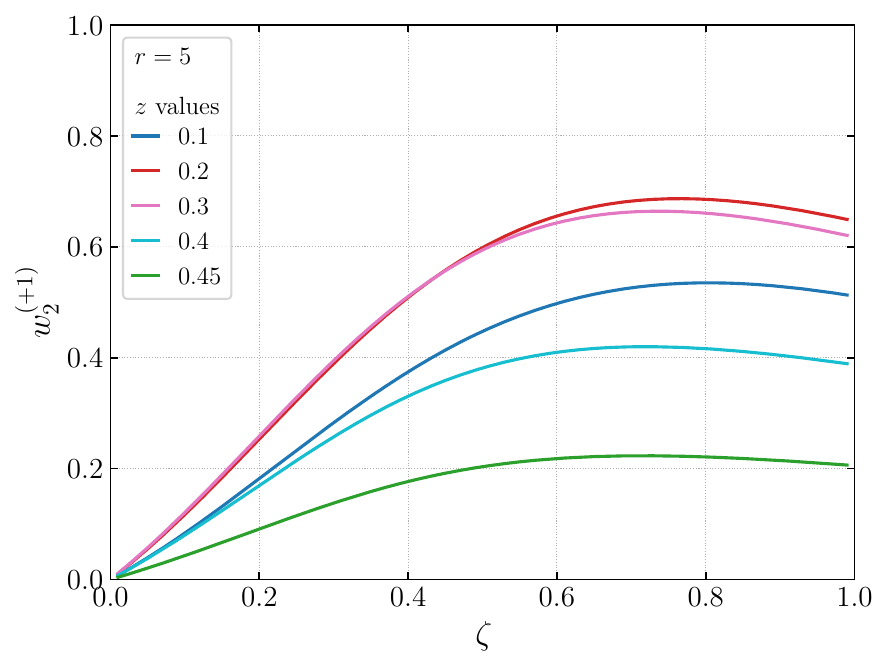}
         \label{fig:three sin x}
     \end{subfigure}
     \begin{subfigure}[h]{0.49\textwidth}
         \centering
         \includegraphics[width=\textwidth]{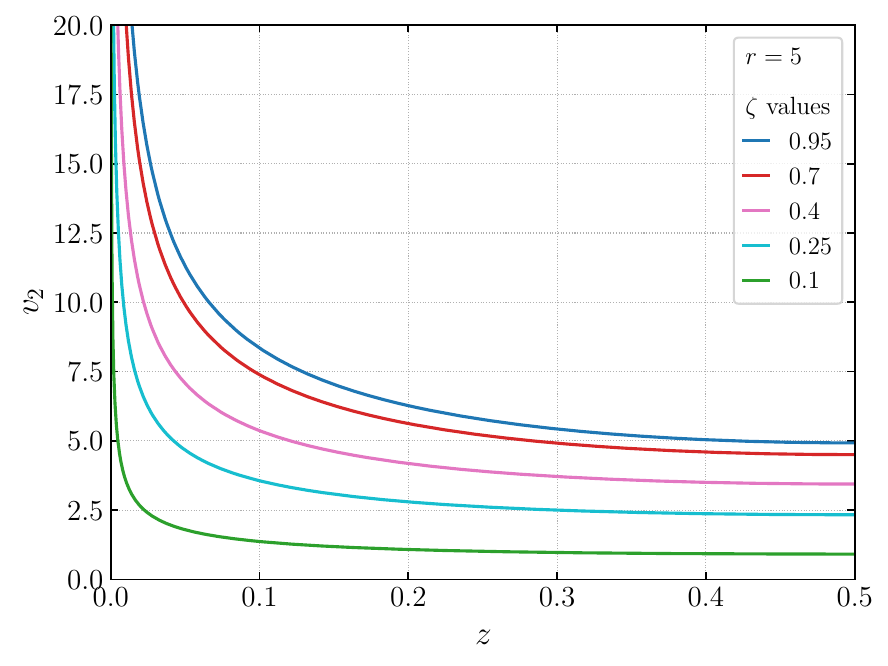}
         \label{fig:three sin x}
     \end{subfigure}
     \begin{subfigure}[h]{0.49\textwidth}
         \centering
         \includegraphics[width=\textwidth]{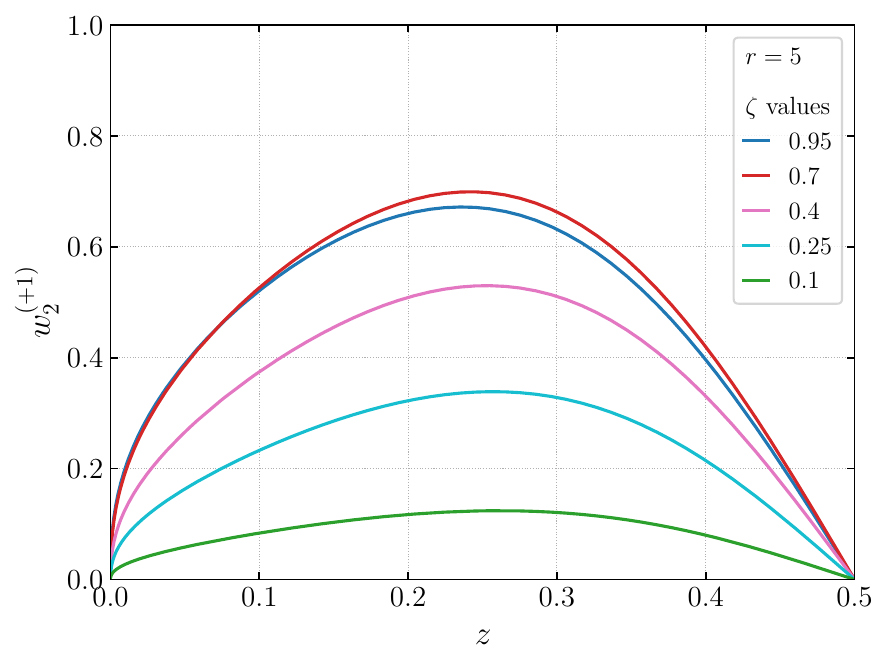}
         \label{fig:three sin x}
     \end{subfigure}
     \caption{\textbf{Left:} Evolution of  $v_2$ in Eq.~\eqref{harmonic_exp} as a function of $\zeta$ (top line) or of $z$ (bottom line) and for different values of $z$ or of $\zeta$, respectively, with fixed $r=5$. \textbf{Right:} equivalent plots for $w_2^{(+1)}$ in Eq.~\eqref{harmonic_exp}. The vacuum piece is subtracted.}
     \label{fig:massless_harm}
\end{figure}

\begin{figure}
     \centering
     \begin{subfigure}[h]{0.49\textwidth}
         \centering
         \includegraphics[width=\textwidth]{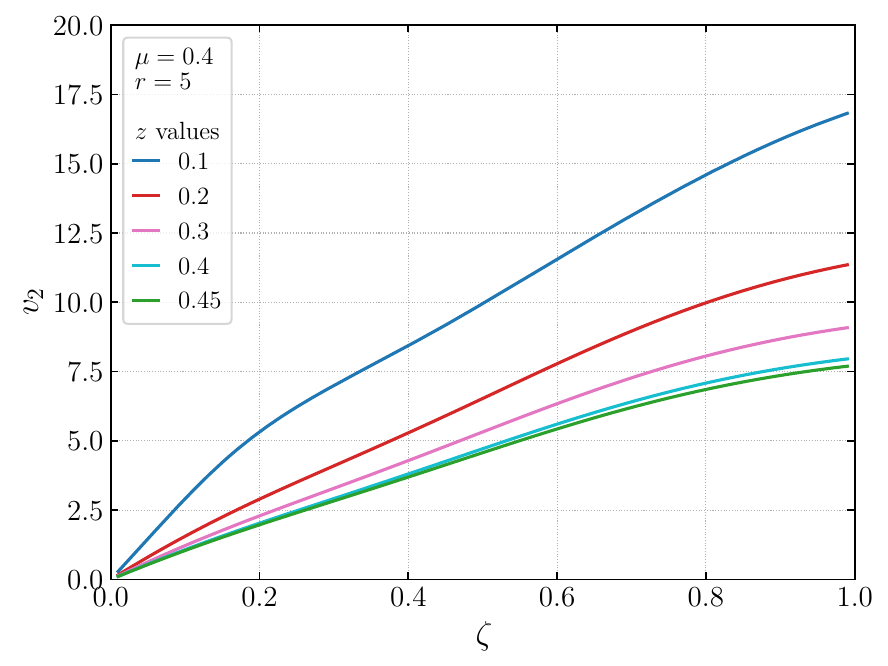}
         \label{fig:three sin x}
     \end{subfigure}
     \begin{subfigure}[h]{0.49\textwidth}
         \centering
         \includegraphics[width=\textwidth]{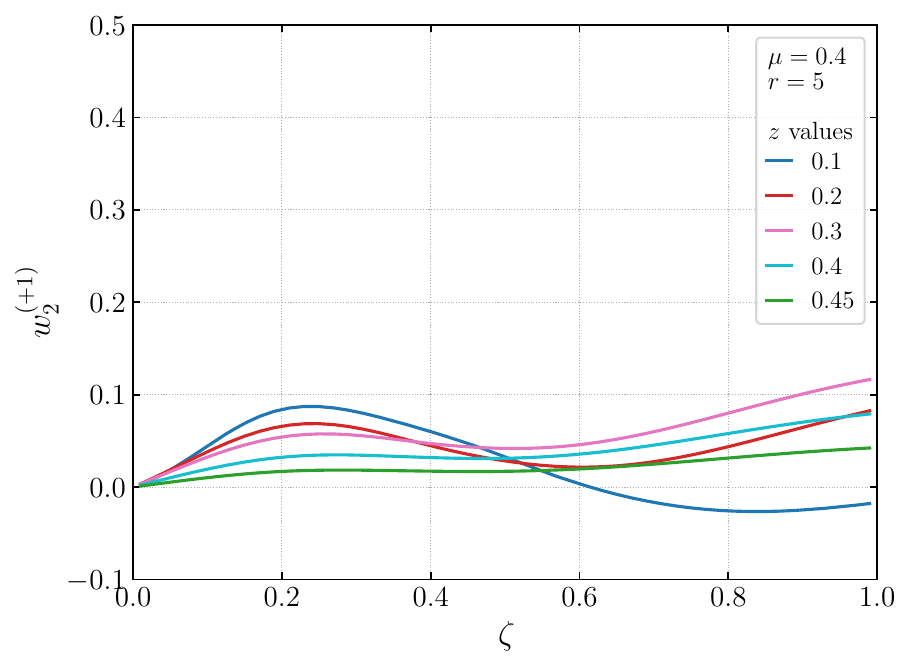}
         \label{fig:three sin x}
     \end{subfigure}
     \begin{subfigure}[h]{0.49\textwidth}
         \centering
         \includegraphics[width=\textwidth]{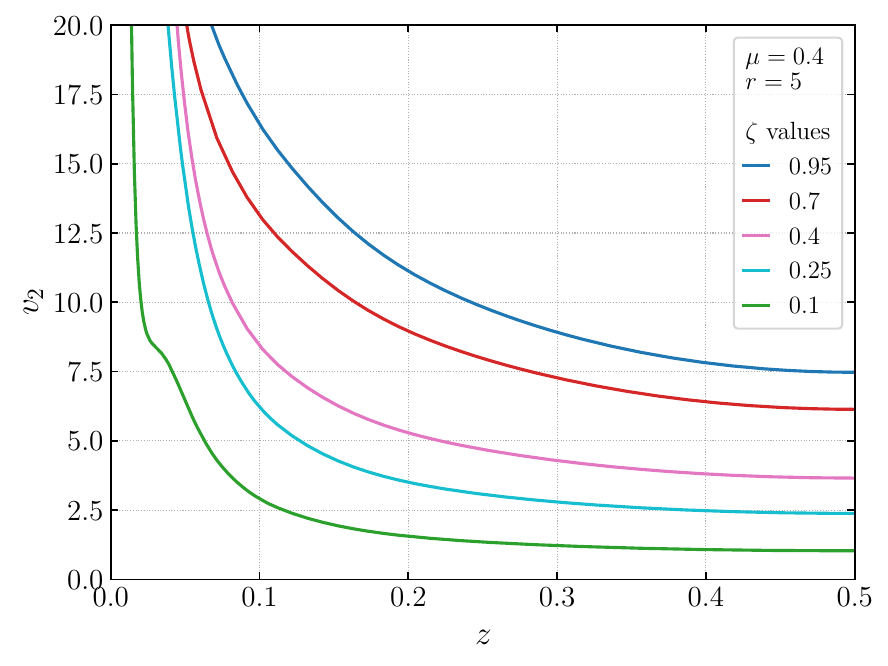}
         \label{fig:three sin x}
     \end{subfigure}
     \begin{subfigure}[h]{0.49\textwidth}
         \centering
         \includegraphics[width=\textwidth]{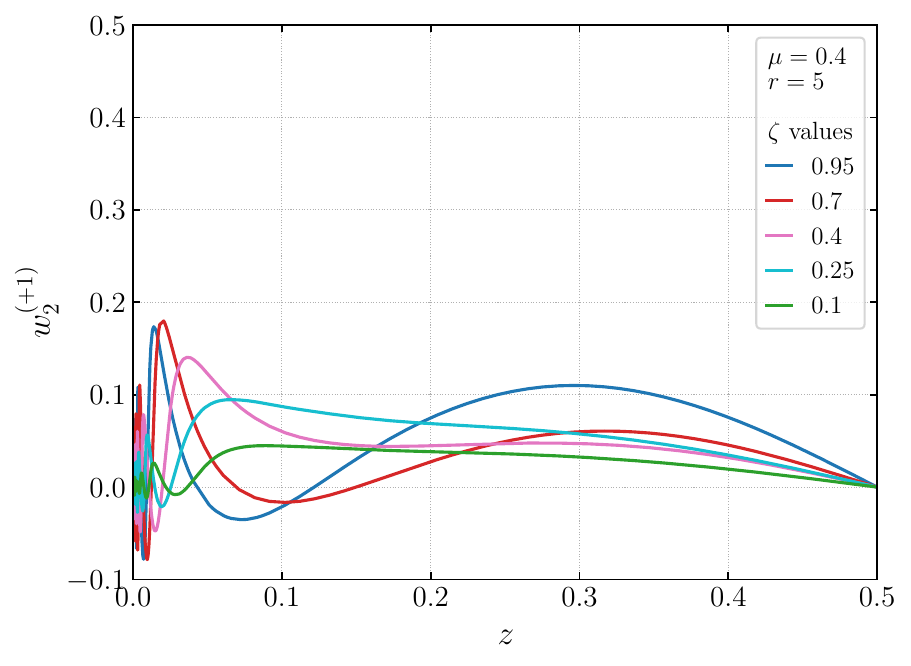}
         \label{fig:three sin x}
     \end{subfigure}
     \caption{\textbf{Left:} Evolution of  $v_2$ in Eq.~\eqref{harmonic_exp} as a function of $\zeta$ (top line) or of $z$ (bottom line) and for different values of $z$ or of $\zeta$, respectively, with fixed $r=5$ and $\mu = 0.4$. \textbf{Right:} equivalent plots for $w_2^{(+1)}$ in Eq.~\eqref{harmonic_exp}. The vacuum piece is subtracted.}
     \label{fig:massive_harm}
\end{figure}

In Figs.~\ref{fig:massless_harm} we show the evolution of the leading harmonics, $v_2$ and $w_2^{(+1)}$, with $\zeta$ (top) and $z$ (bottom). An equivalent plot for massive quarks is shown in Fig.~\ref{fig:massive_harm}. First, when comparing the $v_2$ with $w_2^{(+1)}$ harmonics in absolute value, we get a possible explanation for the near symmetry around $\pi/2$ on the left column plots of Fig.~\ref{azimuthal_plot_phi} and ~\ref{massive_azimuthal_plot_phi}, since this indicates that $\cos(2\phi)$ dominates. Then, $v_2$ is larger for more imbalanced splittings and grows with the absolute value of the anisotropy parameter, as one should expect. In contrast, the evolution of the $w_2$ harmonic is non-monotonic with the energy fraction; this is a consequence of the fact that for $z=0,\, 0.5$ there is no helicity/spin dependence and thus the harmonic has to vanish as shown. As a result, the curves display a maximum for intermediates values of $z$. The evolution in $\zeta$ is also surprising since at large values, the curves tends to flatten and some even display a maximum. These observations are even more striking in the massive case, where the flow coefficient can become negative and it exhibits a more complex evolution in $z$. We note that this behavior can be tied to the larger value of $\mu=0.4$ considered, where some of the assumptions used can break down. Additionally, we see quite smaller values of $w_2$ for the massive case, explaining the less pronounced distinction between spin states on the right column plots of Fig.~\ref{massive_azimuthal_plot_phi}. Overall, mass seems to confound the interesting features of the azimuthal particle distribution.

To summarize, extracting the flow harmonics inside jets would provide a new window into the jet substructure, with a clear sign for non-trivial medium properties being isolated from more traditional medium effects, e.g. momentum broadening, energy loss, which are mainly isotropic and thus can generate any harmonic but $v_0$.

\subsection{Transverse polarization for a massive antenna}
\begin{figure}
    \centering
    \begin{subfigure}[h]{0.49\textwidth}
         \centering
         \includegraphics[width=\textwidth]{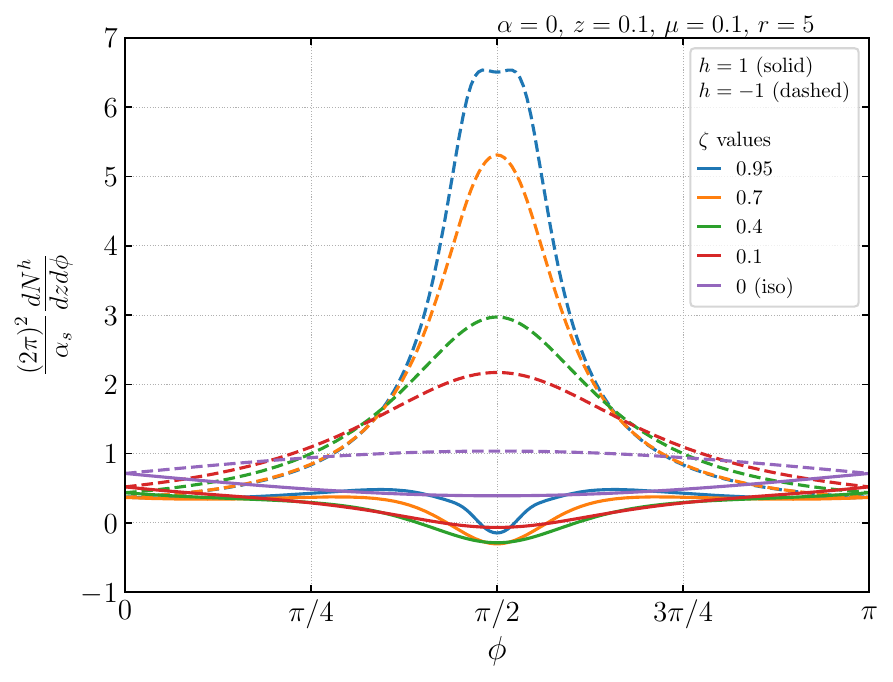}
         \label{fig:three sin x}
     \end{subfigure}
     \begin{subfigure}[h]{0.49\textwidth}
         \centering
         \includegraphics[width=\textwidth]{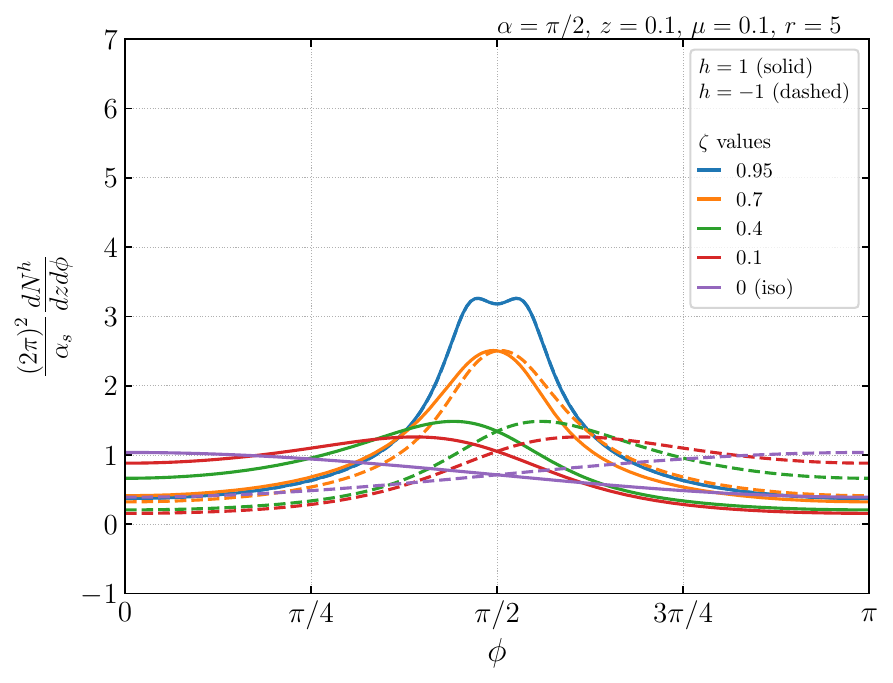}
         \label{fig:three sin x}
     \end{subfigure}
     \begin{subfigure}[h]{0.49\textwidth}
         \centering
         \includegraphics[width=\textwidth]{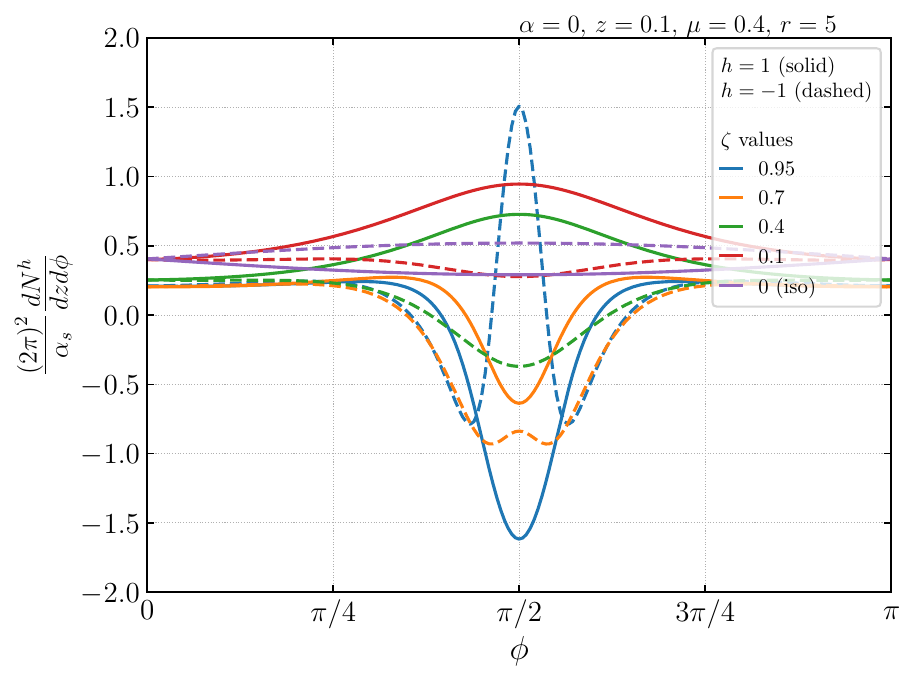}
         \label{fig:three sin x}
     \end{subfigure}
     \begin{subfigure}[h]{0.49\textwidth}
         \centering
         \includegraphics[width=\textwidth]{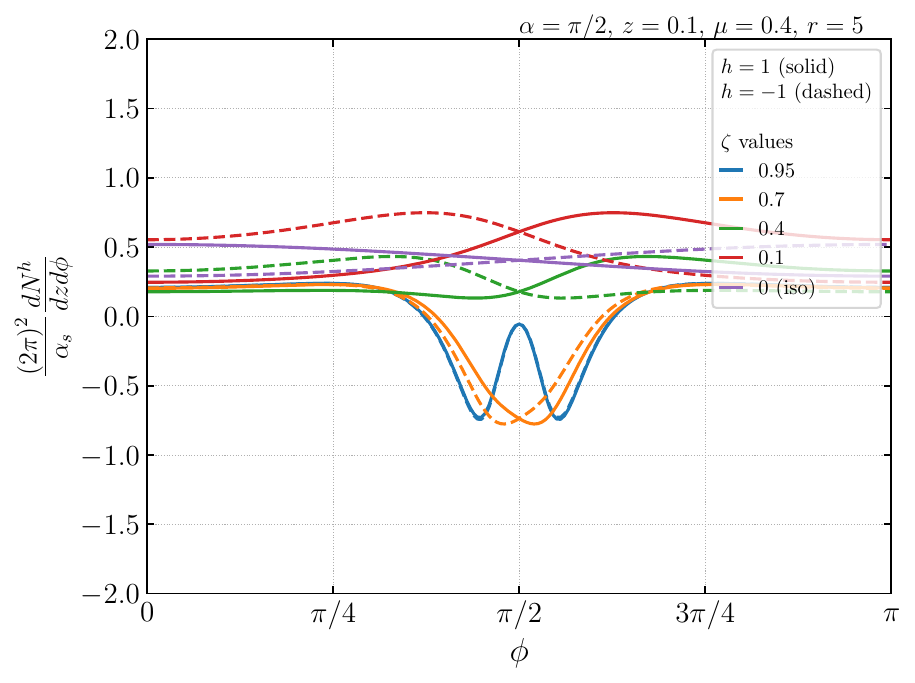}
         \label{fig:three sin x}
     \end{subfigure}
     \caption{Plots of the distribution in Eq.~\eqref{massive_tvaligned_final} as a function of $\phi$ for different values of the anisotropy parameter $\zeta$, for $\alpha=0$ (left) and $\alpha = \pi/2$ (right), for fixed $r=5$, $z=0.1$ and for two different values of $\mu$ -- $0.1$ (upper) and $0.4$ (lower). Note that the spin index $r$ is renamed to $h$. The vacuum piece is subtracted.}
     \label{tvPol_1}
\end{figure}

\begin{figure}
    \centering  
    \begin{subfigure}[h]{0.49\textwidth}
         \centering
         \includegraphics[width=\textwidth]{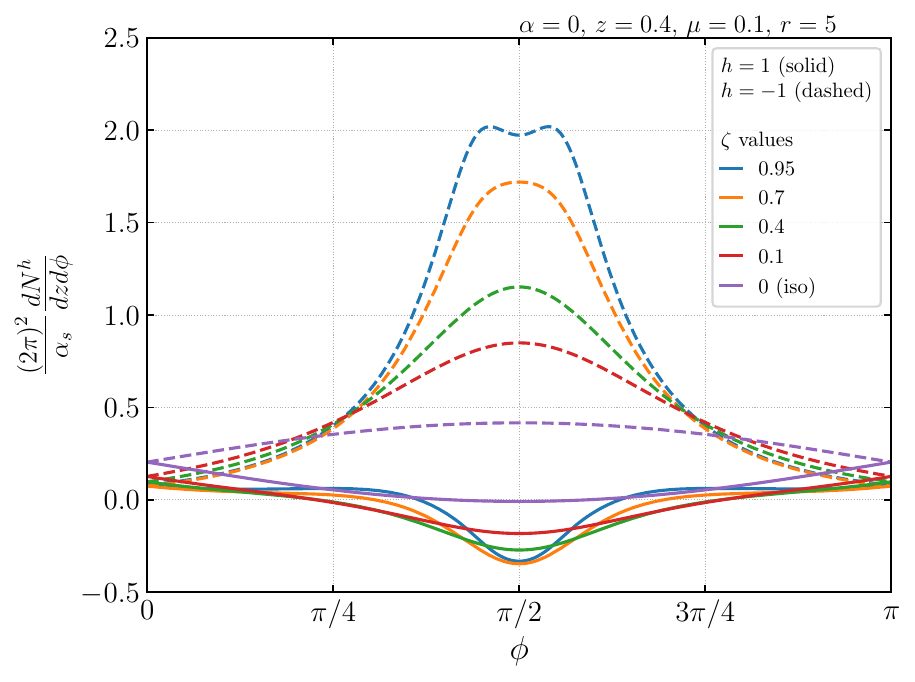}
         \label{fig:three sin x}
     \end{subfigure}
     \begin{subfigure}[h]{0.49\textwidth}
         \centering
         \includegraphics[width=\textwidth]{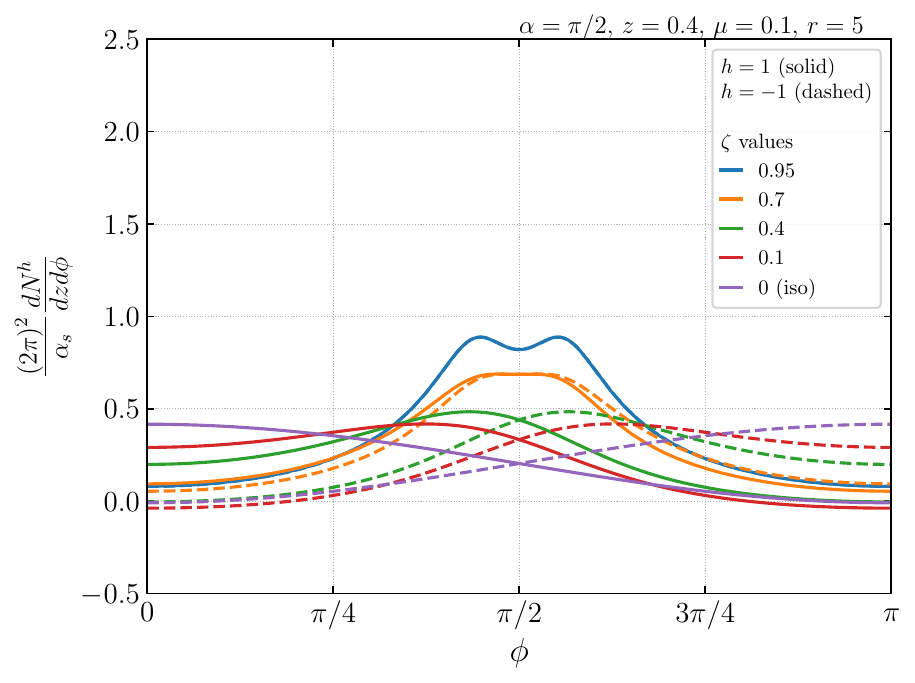}
         \label{fig:three sin x}
     \end{subfigure}
     \begin{subfigure}[h]{0.49\textwidth}
         \centering
         \includegraphics[width=\textwidth]{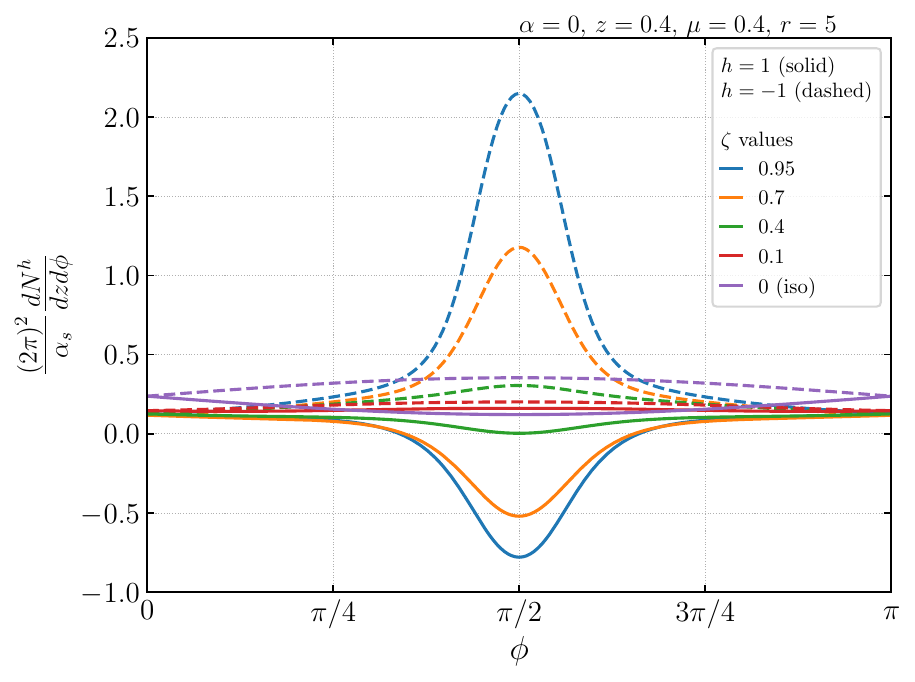}
         \label{fig:three sin x}
     \end{subfigure}
     \begin{subfigure}[h]{0.49\textwidth}
         \centering
         \includegraphics[width=\textwidth]{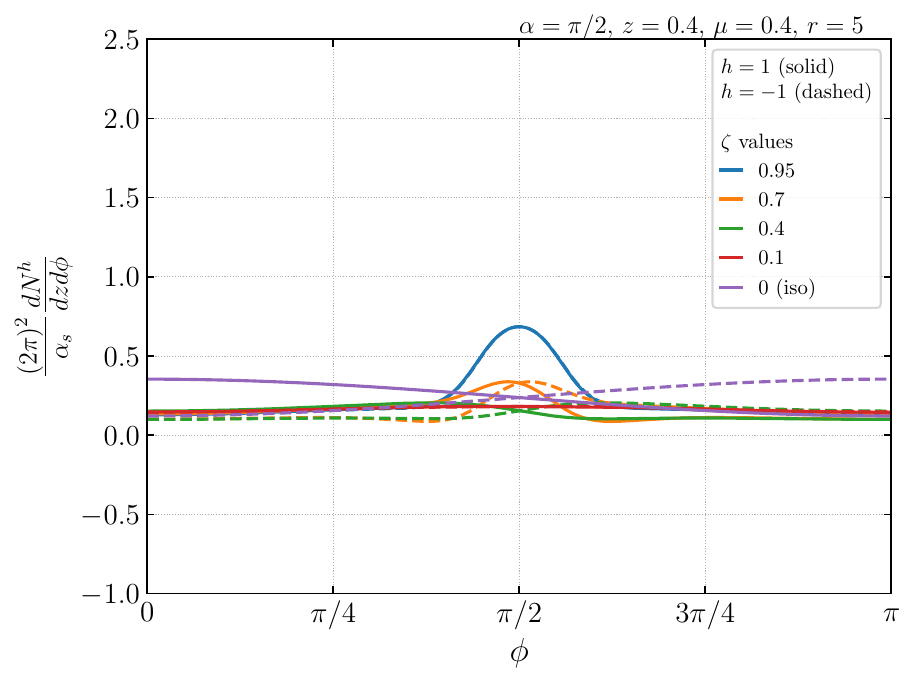}
         \label{fig:three sin x}
     \end{subfigure}
     \caption{Plots of the distribution in Eq.~\eqref{massive_tvaligned_final} as a function of $\phi$ for different values of the anisotropy parameter $\zeta$, for $\alpha=0$ (left) and $\alpha = \pi/2$ (right), for fixed $r=5$, $z=0.4$ and for two different values of $\mu$ -- $0.1$ (upper) and $0.4$ (lower). Note that the spin index $r$ is renamed to $h$. The vacuum piece is subtracted.}
     \label{tvPol_2}
\end{figure}

So far we have defined the quark and anti-quark spin states as the projection along the $z$ axis. However, because we are trying to tap the interplay between spin and the anisotropy the $q\bar q$ antenna experiences in the transverse plane, it might be more interesting to consider spin projected along some axis on that plane. One can do so by changing the expression for the vertex following the steps detailed in Appendix~\ref{App1} and inserting it in Eq.~\eqref{eq:dN_1}. Carrying out the remaining of the calculation exactly as before, the particle distribution reads 
\begin{align}\label{massive_tvaligned_final}
    & \frac{dN^{r s }}{dzd\phi} = 
    \frac{\alpha_s}{4(2 \pi)^2}\Re  \int_{X_v^+} e^{-i\frac{m^2}{2\tilde{q}_0^+}\Delta t} \frac{\sqrt{c_{1x}}\sqrt{c_{1y}}}{\sqrt{c_{3x}}\sqrt{c_{3y}}}\delta_{rs}\nn
    & \left(\frac{(c_{1y}c_{2y}c_{3x}-c_{1x}c_{2x}c_{3y})(c_{3y}\cos^2{\phi}-c_{3x}\sin^2{\phi})}{\left(c_{3y}\cos^2{\phi}+c_{3x}\sin^2{\phi}\right)^2} +\left(\frac{m^2}{2\tilde{q}_0^+} + c_4\right)\frac{c_{3x}c_{3y}}{c_{3y}\cos^2{\phi}+c_{3x}\sin^2{\phi}} + \right.\nn
    & \left. -\frac{(1-i)mr \sqrt{\pi}}{2\sqrt{\tilde{q}_0^+}}\sqrt{\frac{c_{3x}c_{3y}}{c_{3y}\cos^2\phi+c_{3x}\sin^2\phi}}\frac{(2ic_{1x}-c_{2x})c_{3y}\cos\phi\sin\alpha - (2ic_{1y}-c_{2y})c_{3x}\sin\phi\cos\alpha}{\left(c_{3y}\cos^2{\phi}+c_{3x}\sin^2{\phi}\right)}\right) \, ,
\end{align}
where $\alpha$ is the angle that the spin quantization axis makes with the $x$-axis, i.e., $\alpha=0,\pi/2$ corresponds to measuring spin along the $x$ and $y$ axis, respectively. As discussed in Appendix~\ref{App1} and as is clear from the previous expression, when there is a very large anisotropy, $\hat q_y \gg \hat q_x$ ($\zeta \rightarrow 1$), the outgoing $q\bar q$ pair is unpolarized along the $y$ direction ($\alpha = \pi/2$). However, when measuring spin along the $x$-axis ($\alpha=0$), the resulting distribution is polarized. Notice that this effect is solely driven by the mass dependent term. This observation can be confirmed in Fig.~\ref{tvPol_1} and \ref{tvPol_2}, where we plot Eq.~\eqref{massive_tvaligned_final} for several values of $\zeta$, $z$ and $\mu$ as a function of $\phi$. 
Indeed, for very large anisotropies like $\zeta=0.95$ and $0.7$, one can clearly observe that when $\alpha=\pi/2$ (right plots) the solid and dashed curves are quite similar to each other, reflecting a result that is nearly independent of $h$ and thus a $q\bar q$ that is very weakly polarized along the $y$-axis. For $\alpha = 0$, the two curves are significantly different around $\phi=\pi/2$, giving rise to a significant polarization along the $x$-axis. 
Even for intermediate values of anisotropy, for instance $\zeta = 0.4$ (the green curve), the $h$ dependence, i.e., degree of polarization, is quite clearly larger along the $x$-axis than along the $y$-axis. Regarding the dependence in $z$, we see that the softer splittings ($z=0.1$ in Fig.~\ref{tvPol_1} ) seem to give rise to a more pronounced separation between distributions of different spin states, enhancing the polarization effect. The mass $\mu$, despite being the parameter enabling this interesting effect, seems to suppress the degree of polarization when taking larger values (bottom plots of Fig.~\ref{tvPol_1} and \ref{tvPol_2}). Additionally, one should note that in the isotropic case (violet curve in all plots) the $q\bar q$ pair also obtains polarization due to interactions with the medium. Nevertheless, the stark difference with respect to the anisotropic case is that the degree of polarization is the exact same in all directions $\alpha$. This is clear if one plots the distribution as a function of $\phi-\alpha$ since, in the isotropic limit of Eq.~\eqref{massive_tvaligned_final}, the spin-dependent term is simply proportional to $\cos(\phi-\alpha)$. Thus, for anisotropic matter, this direction dependent polarization effect gives yet another feature which can be explored to extract information from the medium, since it informs about the direction of maximal $\hat q$. 

\section{Conclusions}\label{sec:conclusion}
In this work we have studied $q\bar q$ pair production from a gluon in the presence of an anisotropic background QCD medium, as expected to exist in several of the stages characterising the aftermath of URHICs, in particular in the very initial stages, possibly dominated by strong longitudinal color fields. We consider a simple medium model, where the background is static and confined inside a box of fixed longitudinal size. We introduce anisotropic effects by allowing the jet quenching parameter to differ in two orthogonal directions in the transverse plane with respect to the jet axis. This leads to the final particle distribution evolving differently in each of these directions. Furthermore, we keep track of the helicity/spin dependence of the final states, which we show directly couples to the medium anisotropy, offering a new window to study directional effects inside jets in heavy ions collisions. This possibility is very appealing as the direction of the anisotropy is easy to identify experimentally. The fact that spin flip is energy suppressed during the propagation of the parton indicates that these spin-effects from the very early times could survive the subsequent evolution of the medium.

Our final main results, given in Eqs.~\eqref{azimuth_opening_angle_final} and \eqref{massive_zaligned_final}, can be straightforwardly extended beyond the approximations we consider here, although we do not expect the main physical picture portrayed to change qualitatively. Further, we discuss their potential impact in several jet observables, indicating when clear signatures of the medium anisotropy can be found in the jet distribution. As examples of these, we propose the study of Fourier azimuthal harmonics inside
jets as a way to probe non-trivial correlations. Lastly, we argue that spin measurements in different directions allow to probe the polarization of partons inside the jet, further testing directional effects induced by the medium. We leave a more detailed phenomenological study of such observables for future work.

Finally, we note that the presented theoretical results can, in principle, be extended to other hard probes, such as quarkonia or heavy flavour. We believe that this would allow to expand the current toolkit of tomographic tools of the matter produced in URHICs, allowing for a more complete energy scan, which is not possible by just using high energy jets, see e.g.~\cite{Kumar:2022ylt,Du:2023izb}.

\acknowledgments
We wish to thank A. V. Sadofyev and R. Szafron for helpful discussions. We are especially thankful to A. V. Sadofyev for comments concerning the Fourier decomposition of the particle distribution. This work is supported by the European Research Council project ERC-2018-ADG-835105 YoctoLHC; by Maria de Maeztu excellence unit grant CEX2023-001318-M and project PID2020-119632GB-I00 funded by MICIU/AEI/10.13039/501100011033; by ERDF/EU; and by Fundação para a Ciência e a Tecnologia (FCT, Portugal) under project CERN/FIS-PAR/0032/2021. It has received funding from Xunta de Galicia (CIGUS Network of Research Centres); and from the European Union’s Horizon 2020 research and innovation programme under grant agreement No. 824093. The Feynman diagrams in this work were produced using the {\tt JaxoDraw} software~\cite{Binosi:2003yf}. JB is supported by the United States Department of Energy under Grant Contract DESC0012704. JMS is supported by Fundação para a Ciência e Tecnologia (Portugal, FCT) under contract PRT/BD/152262/2021.

\newpage
\appendix
\section{Effective in-medium Feynman rules in coordinate space}\label{App1}
In this appendix we briefly review some elements entering the main text's calculation. Further details can be found in e.g. ~\cite{Mehtar-Tani:2013pia,Blaizot:2012fh}.
\subsection{Parton scalar propagators}
The quark propagator in the presence of a classical background gauge field is given at leading order in powers of $1/p^+$ by
\begin{equation}
    \Gq{ij}{y^+}{y}{x^+}{x}{p^+} = \int^{\boldsymbol{r}(y^+) = \boldsymbol{y}}_{\boldsymbol{r}(x^+)=\boldsymbol{x}}\mathcal{D}\boldsymbol{r}(\xi)\Exp{i\frac{p^+}{2}\int_{x^+}^{y^+}ds^+ \, \dot{\boldsymbol{r}}^2}U^{ij}(x^+,y^+,\boldsymbol{r}(\xi))\, ,
\end{equation}
where the fundamental Wilson line is given by\\
\begin{equation}
    U^{ij}(x^+,y^+,\boldsymbol{r}) = \mathcal{P}\Exp{ig\int_{x^+}^{y^+}ds^+\, \cA_a^{-}(s^+, \boldsymbol{r}(s^+))t^a_{ij}}\, .
\end{equation}
The anti-quark propagator $\overline{\mathcal{G}}$ is the same as the quark propagator with the replacement $U\rightarrow U^{\dagger}$. As for the gluon, one only needs to replace the fundamental Wilson line by an adjoint one $U_A$,\\
\begin{equation}
    U_A^{ab}(x^+,y^+,\boldsymbol{r}) = \mathcal{P}\Exp{g\int_{x^+}^{y^+}ds^+ \, \cA_c^{-}(s^+, \boldsymbol{r}(s^+))f^{cab}}\, .
\end{equation}
In practice, one can combine these effective propagators with standard Feynman diagrammatic techniques with the following modifications compared to the canonical rules for QCD:
\begin{equation}\label{propagator_lines}
    \begin{split}
        \text{\textbf{internal line}} \qquad \rightarrow \qquad & \Gq{ij}{y^+}{y}{x^+}{x}{p^+}/2p^+\\[2ex]
        \text{\textbf{incoming external line}} \qquad \rightarrow \qquad & \Gq{ij}{y^+}{y}{x^+}{x}{p^+} e^{-i(\frac{\boldsymbol{p}^2}{2p^+}x^+-\boldsymbol{p}\cdot \boldsymbol{x})}\\[2ex]
        \text{\textbf{outgoing external line}} \qquad \rightarrow \qquad & \Gq{ij}{y^+}{y}{x^+}{x}{p^+} e^{i(\frac{\boldsymbol{p}^2}{2p^+}y^+-\boldsymbol{p}\cdot \boldsymbol{y})}\\
    \end{split}    
\end{equation}
For massive quarks, one only needs to account for the additional phases coming from the intermediate integrations over $p_n^-$ in between each medium scattering, all of which cancel except those at the light-cone time limits, i.e., one makes the replacement:
\begin{equation}\label{massive_phase}
        \Gq{ij}{y^+}{y}{x^+}{x}{p^+}_{m\neq 0}\rightarrow \Gq{ij}{y^+}{y}{x^+}{x}{p^+}_{m= 0} \Exp{-i\frac{m^2}{2p^+}(y^+-x^+)}\, ,
\end{equation}
where $m$ is the quark mass.
~\\
\subsection{Spin/helicity-dependent quark-gluon vertex}
\indent We assume the quark-gluon vertex is unchanged by the presence of the medium and we write it as
\begin{equation}\label{general_vertex}
    (A^{\lambda r s})_{kl}^b(z,\boldsymbol{P}) = ig t_{kl}^b V^{\lambda r s}(z,\boldsymbol{P}) \, ,
\end{equation}
where the relative transverse momentum reads $\boldsymbol{P} = \boldsymbol{p}_q-z\boldsymbol{p}_g$, with $z$ the fraction of $+$ momentum from the gluon carried by the quark, $\boldsymbol{p}_g$ the former's transverse momentum and $\boldsymbol{p}_q$ the latter's. At leading order in $1/p^+$, the only transverse momentum dependence of the vertex comes from a linear term in $\P$, as we will see below. The labels $\lambda, r,s$ represent, respectively, the gluon polarization state in circular basis and the $q$ and $\bar q$ spin projection along a general spin quantization axis or the helicity states.\\
\indent To insert a momentum-space vertex via coordinate space rules, one replaces transverse momenta by derivatives of $\delta$-functions. This can be explained by going from the mixed space (transverse momentum + light-cone time) formulation of the amplitude to the coordinate space one. The amplitude in Fig.~\ref{fig:diagram} would have the following form in mixed space:\\
\begin{equation}
    i\mathcal{M} = \int_{\boldsymbol{p}_q,...}\boldsymbol{p}_q\text{ }\Gq{ik}{L^+}{p_1}{x_v^+}{p_q}{p_1^+} R^{kj}(\boldsymbol{p}_q,...)
\end{equation}
where $R^{kj}(\boldsymbol{p}_q,...)$ represents the remaining part of the amplitude. Let us focus only on the vertex term proportional to $\p_q$ (the other term in $\P$ is completely analogous). We would like to write the vertex in such a way that one could directly extract the coordinate space Feynman rule for it. To do so, we just need to use the convolution theorem for Fourier transforms:\\
\begin{equation}\label{convolution_vertex}
    \begin{split}
    i\mathcal{M} & = \int_{\boldsymbol{p}_q,...} \mathcal{F}\left(\mathcal{F}^{-1}(\boldsymbol{p}_q)\ast \mathcal{F}^{-1}(\Gq{ik}{L^+}{p_1}{x_v^+}{p_q}{p_1^+})\right)(\boldsymbol{p}_q)R^{kj}(\boldsymbol{p}_q,...)\\
    & = \int_{\boldsymbol{p}_q,...} \int_{\boldsymbol{x_v}}e^{i\boldsymbol{x_v}\cdot \boldsymbol{p}_q}\int_{\boldsymbol{y}} i\delta^{'}(\boldsymbol{y}) \Gq{ik}{L^+}{\boldsymbol{p_1}}{x_v^+}{\boldsymbol{x_v-y}}{p_1^+} R^{kj}(\boldsymbol{p}_q,...)\\
    & = \int_{\boldsymbol{x_v},...} \int_{\boldsymbol{y}} i\delta^{'}(\boldsymbol{y}) \Gq{ik}{L^+}{\boldsymbol{p_1}}{x_v^+}{\boldsymbol{x_v-y}}{p_1^+} R^{kj}(\boldsymbol{x_v},...)\, .
    \end{split}
\end{equation}
Hence, to write the vertex directly in coordinate space one should replace the momentum coming out of the vertex as $\boldsymbol{p}\rightarrow i\delta^{'}(\boldsymbol{y})$ and displace the vertex coordinate associated to the corresponding scalar propagator by $-\boldsymbol{y}$, integrating over this displacement. For momenta coming into the vertex the same reasoning holds but the correct momentum replacement is $\boldsymbol{q}\rightarrow -i\delta^{'}(\boldsymbol{y})$. Accordingly, to write the vertex in Eq.~\eqref{general_vertex} in coordinate space, one attaches coordinate displacement $\x$ to the incoming gluon and $\y$ to the outgoing quark and makes the substitution:
\begin{equation}\label{momentum_substitution}
    \P \rightarrow i(\delta(\x)\delta'(\y)+z\delta'(\x)\delta(\y)) \, .
\end{equation}
Further below, when presenting the squared expression for the vertex, we will use $\overline{\P}$ for the relative momentum in the complex conjugate amplitude. One should keep in mind the correct substitution for this momentum is the complex conjugate of Eq.~\eqref{momentum_substitution} with different (dummy) coordinate displacements.

The helicity-dependent vertex for circularly polarized gluons and for massless quarks is given by
\begin{equation}\label{massless_vertex}
    \begin{split}
    & V^{\lambda h h'}(z,\boldsymbol{P}) = 2\gamma^{\lambda h}(z)\delta_{h,-h'}(\boldsymbol{\epsilon}^{\lambda}\cdot \boldsymbol{P})\\
    & \gamma^{\lambda h}(z) = \frac{1}{\sqrt{z(1-z)}}\left(z\delta^{\lambda h} - (1-z)\delta^{\lambda -h}\right) \, .
    \end{split}
\end{equation}
As stated in the main text, the quark's helicity is given by $h=\pm 1$, the anti-quark's by $-h$ and the gluon polarization vector by $\boldsymbol{\epsilon}^{\lambda} = \frac{1}{\sqrt{2}}(1,\lambda i)$, i.e, the circular basis with $\lambda =\pm 1$. When squaring and averaging over polarizations one gets:\\
\begin{equation}\label{massless_squared}
    \begin{split}
    & \frac{1}{2}\sum_{\lambda}(V^{\lambda h h'})(z,\boldsymbol{P})\left((V^{\lambda h h'})(z,\boldsymbol{\overline{P}})\right)^{*} =\\
    & \delta_{h,-h'}\frac{2}{z(1-z)} \left(P_{qg}^{vac}(z)(\boldsymbol{P}\cdot\boldsymbol{\overline{P}}) + \frac{ih}{2}(1-2z)(\P \times \overline{\P})_z\right) \, ,
    \end{split}
\end{equation}
where the $g\rightarrow q\bar q$ vacuum splitting function reads $P_{qg}^{vac}(z) = T_R(z^2+ (1-z)^2)$ and $(\P \times \overline{\P})_z = \P_x\overline{\P}_y-\P_y\overline{\P}_x$.

For massive quarks, a (non-suppressed) term where the quark and anti-quark spins can be anti-parallel arises , i.e.,
\begin{equation}\label{massive}
    V^{\lambda r s}(z,\boldsymbol{P}) =  2r\,\gamma^{\lambda r}(z)\delta_{rs} (\boldsymbol{\epsilon}^{\lambda}\cdot \boldsymbol{P}) - \delta^{\lambda r}\delta_{r,-s} \frac{2m}{\sqrt{2z(1-z)}}\, ,
\end{equation}
where now we work with spin states $r,s$ instead of helicity states since the latter ones are not Lorentz invariant for massive quarks.  The parallel spins term also contains mass dependent terms, but these are suppressed by an energy denominator with respect to the remaining. When squaring and averaging over polarizations one gets:
\begin{equation}\label{massive_squared}
    \begin{split}
    & \frac{1}{2}\sum_{\lambda}(V^{\lambda rs})(z,\boldsymbol{P}) \left((V^{\lambda rs})(z,\boldsymbol{\overline{P}})\right)^{*} = \\
    & \frac{2}{z(1-z)} \left\{\delta_{rs}\left(P_{qg}^{vac}(z)(\boldsymbol{P}\cdot\boldsymbol{\overline{P}}) + \frac{ir}{2}(1-2z)(\P \times \overline{\P})_z\right) + \delta_{r,-s}\frac{m^2}{2}\right\} \, .
    \end{split}
\end{equation}

It can also be interesting to measure the spins of the $q\bar q$ pair along some particular axis, in which case one can generalize the previously obtained results to a general spin-quantization axis $\Vec{n} = \left(\sin\theta\cos\alpha, \sin\theta\sin\alpha, \cos\theta\right)$ ($\theta = 0, \alpha = \pi$ gives us spin quantized along the $z$ axis, as in Eq.~\eqref{massive}):
\begin{equation}\label{gen_spin_circg}
    \begin{split}
        & V^{\lambda r s}(z,\boldsymbol{P}) = \frac{2}{\sqrt{z(1-z)}}\\
        & \left\{ \left. r\delta_{r s}\left(\left[ \cos\theta\left(z\delta^{\lambda r} - (1-z)\delta^{\lambda,-r}\right) + \sin^2(\theta/2)\left(\delta^{\lambda r} - \delta^{\lambda, -r}\right) \right] (\boldsymbol{\epsilon}^{\lambda}\cdot\boldsymbol{P}) - \frac{m}{2\sqrt{2}}\sin\theta e^{i\lambda\alpha}\right)+\right. \right.\\
        & \left. + \delta_{r,-s}\left(\sin\theta(z-1/2)(\boldsymbol{\epsilon}^{\lambda}\cdot\boldsymbol{P}) + \frac{m}{\sqrt{2}}e^{i\lambda\alpha}\left[\delta^{\lambda r}\cos^2(\theta/2) - \delta^{\lambda, -r} \sin^2(\theta/2)\right]\right)\right\} \, .
    \end{split}
\end{equation}
Furthermore, if we consider $\theta \neq 0$, i.e, a spin-quantization axis different from $z$, cross terms which are linear in the momenta (derivatives in the coordinate space amplitude) and mass appear when squaring the vertex. For instance, if we set $\theta=\pi/2$, i.e., measuring the spin of the quark and anti-quark along some axis in the transverse plane, square Eq.~\eqref{gen_spin_circg} and average over polarization states we get:
\begin{equation}\label{massive_tvpol_vertex}
    \begin{split}
    & \frac{1}{2}\sum_{\lambda}(V^{\lambda r s})(z,\boldsymbol{P})\left((V^{\lambda r s})(z,\boldsymbol{\overline{P}})\right)^{*} = \\
    & \frac{1}{2z(1-z)}\left\{(\boldsymbol{P}\cdot\boldsymbol{\overline{P}})\left(\delta_{rs} + (2z-1)^2\delta_{r,-s}\right) +m^2 \right.\\
    & \left. +imr\left[\left(\boldsymbol{P}^x-\boldsymbol{\overline{P}}^x\right)\sin\alpha - \left(\boldsymbol{P}^y-\boldsymbol{\overline{P}}^y\right)\cos\alpha\right]\left(\delta_{rs} - (2z-1)\delta_{r,-s}\right)\right\} \, .
    \end{split}
\end{equation}
The linear terms vanish in the vacuum limit since the momenta in the direct and complex conjugate amplitudes are the same. This means that the difference $\P^i - \bar \P^i$ is controlled by $\hat q_i$ in the corresponding direction. If $\hat{q}^y \gg \hat{q}^x$, for instance, then the spin dependent term is dominated by the $\cos\alpha$ term. This implies that when measuring spin states along the $x$ axis ($\alpha=0$) the difference between the spectrum for up and down ($r=\pm 1$) is much greater than the difference obtained when measuring spin along the $y$ axis ($\alpha = \pi/2$). Put simply, in the limit where $\hat{q}^x \rightarrow 0$ and $\hat{q}^y$ is finite, then the $q\overline{q}$ pair is completely unpolarized in the $y$ direction, since there is no spin dependence, and polarized in the $x$ direction. The degree of polarization depends on the quark mass and naturally grows with $\hat{q}^y$. 
 
\section{Calculation details}\label{app:App2}
In this appendix we provide further details on the treatment of the color structure and medium averaging underlying the main calculation.

\subsection{Color structure}\label{app:2:color_struct}
Let us explicitly write the Wilson line correlators appearing in Eq.~\eqref{T_expanded} without prefactors and path integrals (to each $r_i$ corresponds a different path integral and the numbering is irrelevant):
\begin{equation}\label{T_color}
    \begin{split}
    & T^{\text{color}} = \frac{1}{N_c^2-1} \left\langle U_A^{ba}(\r_5)(U_A^{\dagger})^{a\overline{b}_1}(\r_9)\right\rangle\\
    & \times t_{kl}^b t_{\overline{l}\overline{k}}^{\overline{b}} \left\langle (U_A^{\dagger})^{\overline{b}_1\overline{b}}(\r_8) (U^{\dagger)})^{lj_1}(\r_4)U^{i_1k}(\r_3)\right\rangle\\ 
    & \times \left\langle U^{ii_1}(\r_1) (U^{\dagger})^{\overline{k}i}(\r_6) (U^{\dagger})^{j_1j}(\r_2)U^{j\overline{l}}(\r_7)\right\rangle \, . \\
    \end{split}
\end{equation}
First note that this object is in a color singlet state. This is because we are considering the final-state quark and anti-quark to have definite color states which are then summed over when squaring the amplitude. Since color is conserved throughout the process (the system as a whole does not exchange color with the medium due to the diagonal color matrix in Eq.~\eqref{pair_correlator}), this means the system is in a singlet state at all times. In particular, after expanding the color structure inside each time region as a linear combination of the state projectors in each of its irreducible representations, one can drop the color non-singlet part (one can include the non-singlet part in the expansion, but the contraction with the final region will always only pick out the singlet parts). Finding out the color singlet states amounts to finding out states for which the second-order Casimir operator in the appropriate representation vanishes \cite{Fukushima:2007dy, Fukushima:2017mko} . For the two-point function the only option is the normalized projector $\frac{\delta^{b\bar b_1}}{\sqrt{N_c^2-1}}$. Hence, we have\\
\begin{equation}
    \left\langle U_A^{ba}(\boldsymbol{r}_5)(U_A^{\dagger})^{a\bar b_1}(\boldsymbol{r}_9)\right\rangle = \frac{\delta^{b\bar b_1}}{N_c^2-1}\left\langle \text{Tr}\left[U_A(\boldsymbol{r}_5)U_A^{\dagger}(\boldsymbol{r}_9)\right]\right\rangle \, .
\end{equation}
Propagating the $\delta^{b\bar b_1}$ to the rest of the squared amplitude, writing the adjoint Wilson line in terms of fundamental ones and repeatedly making use of the Fierz identity, i.e, using 
\begin{align}\label{fierz}
     & U_A^{b\overline{b}} = 2\text{Tr}\left[t^bU^{\dagger}t^{\overline{b}}U\right]\, ,\quad  t^b_{kl}t^b{ij} = \frac{1}{2}(\delta^{kj}\delta^{li}-\frac{1}{N_c}\delta^{kl}\delta^{ij}) \, ,
\end{align}
the other two regions read:\\
\begin{equation}\label{before_proj}
    \begin{split}
    & \frac{1}{2}\Bigg(\left\langle \left(U(\r_3) U^{\dagger}(\r_8)\right)^{i_1 \bar k}\left(U(\r_8)U^{\dagger}(\r_4))\right)^{\bar l j_1}\right\rangle_{(x_v^+,\bar x_v^+)} \left\langle\left( U^{\dagger}(\r_6)U(\r_1)\right)^{\bar k i_1}\left(U^{\dagger}(\r_2)U(\r_7))\right)^{j_1 \bar l}\right\rangle_{(\bar x_v^+, L^+)}  \\ 
    & - \frac{1}{N_c^2}\left\langle\text{Tr}\left[U(\r_3)U^{\dagger}(\r_4)\right]\right\rangle_{(x_v^+, \bar x_v^+)}\left\langle\text{Tr}\left[U^{\dagger}(\r_6)U(\r_1)U^{\dagger}(\r_2)U(\r_7)\right]\right\rangle_{( \bar x_v^+, L+)} \Bigg)  \, .
    \end{split}
\end{equation}
To fully contract the remaining color structure in the first line, one needs to expand each time region's contribution in a basis of color singlets, as we have discussed. For both cases the number of such states is two and they can be obtained through the contractions $\delta^{i_1\bar k}\delta^{\bar l j_1}$ and $\delta^{i_1 j_1}\delta^{\bar l \bar k}$. An orthonormal basis for these singlet projectors is straightforward to obtain: \cite{Fukushima:2007dy, Fukushima:2017mko}:\\
\begin{equation}
    \begin{split}
        & s_1^{i_1\bar k \bar l j_1} = \frac{1}{N_c}\delta^{i_1\bar k}\delta^{\bar l j_1}\\
        & s_2^{i_1\bar k \bar l j_1} = \frac{1}{\sqrt{N_c^2-1}}\left(\delta^{i_1 j_1}\delta^{\bar l \bar k} - \frac{1}{N_c}\delta^{i_1\bar k}\delta^{\bar l j_1}\right) \, .
    \end{split}
\end{equation}
After expanding each correlator in the first line of Eq.~\eqref{before_proj} as a linear combination of the projectors $s_1$ and $s_2$, the coefficients are easily obtained by taking advantage of orthonormality. For instance, we have\\
\begin{equation}
    \begin{split}
    & \left\langle\left( U^{\dagger}(\r_6)U(\r_1)\right)^{\bar k i_1}\left(U^{\dagger}(\r_2)U(\r_7))\right)^{j_1 \bar l}\right\rangle = \frac{1}{N_c}\left\langle\text{Tr}\left[U^{\dagger}(\r_6)U(\r_1)\right]\text{Tr}\left[U^{\dagger}(\r_2)U(\r_7)\right]\right\rangle s_1^{i_1\bar k \bar l j_1}\\
    & + \frac{1}{\sqrt{N_c^2-1}}\left(\left\langle\text{Tr}\left[U^{\dagger}(\r_6)U(\r_1)U^{\dagger}(\r_2)U(\r_7)\right]\right\rangle - \frac{1}{N_c}\text{Tr}\left[U^{\dagger}(\r_6)U(\r_1)\right]\text{Tr}\left[U^{\dagger}(\r_2)U(\r_7)\right] \right)s_2^{i_1\bar k \bar l j_1}
    \end{split}
\end{equation}
and for the remaining correlator it is completely analogous. Using these expansions, the fully contracted result of the color part of the squared amplitude in Eq.~\eqref{T_color} finally simplifies to\\
\begin{equation}
    T^{\text{color}} = \mathcal{C}^{(2)}\mathcal{C}^{(3)}\mathcal{C}^{(4)} \, ,
\end{equation}
where\\
\begin{equation}\label{color_regions}
    \begin{split}
    & \mathcal{C}^{(2)} = \frac{1}{N_c^2 -1}\left\langle \text{Tr}\left[U_A(\boldsymbol{r}_5)U_A^{\dagger}(\boldsymbol{r}_9)\right]\right\rangle\\
    & \mathcal{C}^{(3)} = \frac{1}{2(N_c^2-1)}\left\langle \text{Tr}\left[U^{\dagger}(\r_8)U(\r_3)\right]\text{Tr}\left[U^{\dagger}(\r_4)U(\r_8)\right]  - \frac{1}{N_c}\text{Tr}\left[U^{\dagger}(\r_4)U(\r_3)\right]\right\rangle\\
    & \mathcal{C}^{(4)} = \frac{1}{N_c^2-1}\left\langle \text{Tr}\left[U^{\dagger}(\r_6)U(\r_1)\right]\text{Tr}\left[U^{\dagger}(\r_2)U(\r_7)\right] - \frac{1}{N_c}\text{Tr}\left[U^{\dagger}(\r_6)U(\r_1)U^{\dagger}(\r_2)U(\r_7)\right]  \right\rangle \, ,
    \end{split}
\end{equation}
which resembles results in, e.g., \cite{Apolinario:2014csa, Isaksen:2020npj,Dominguez:2012ad}. Both $\mathcal{C}^{(2)}$ and $\mathcal{C}^{(3)}$ have exact analytical solutions, which are presented in Eq.~\eqref{C2_result} and \eqref{3point_corr}, while $\mathcal{C}^{(4)}$ can only be obtained numerically \cite{Isaksen:2020npj}. In the large-$N_c$ limit, it has been argued \cite{Apolinario:2014csa, Blaizot:2012fh} that the quadrupole (the second term in $\mathcal{C}^{(4)}$) can be approximated by the product of two independent averages of two Wilson line traces, an approximation which is kinematically favoured when the formation time of the antenna is small compared to the medium's length. Nevertheless, in our case, this term is color suppressed by a power $N_c^2$, so we may drop it. This simplification would not be possible in, for instance, the in-medium $\gamma \rightarrow q\bar q$ splitting, where the quadrupole must be dealt with it \cite{Dominguez:2019ges}. As for the first term in $\mathcal{C}^{(4)}$, it was shown in \cite{Isaksen:2020npj} that in the large-$N_c$ limit the differential equation for the average of two traces (first term in $\mathcal{C}^{(4)}$) decouples from the equation for the average of a single trace (second term in $\mathcal{C}^{(4)}$). This allows the former to be calculated exactly in this limit:
\begin{align}
    \left\langle\text{Tr}\left[U^{\dagger}(\r_6)U(\r_1)\right]\text{Tr}\left[U^{\dagger}(\r_2)U(\r_7)\right]\right\rangle &= \left\langle\text{Tr}\left[U^{\dagger}(\r_6)U(\r_1)\right]\right\rangle\left\langle\text{Tr}\left[U^{\dagger}(\r_2)U(\r_7)\right]\right\rangle \nn 
    &+\mathcal{O}\left(\frac{1}{N_c}\right)
\end{align}
Physically, we are approximating the four body average as the independent broadening of the quark and anti-quark.\\

\subsection{2-point functions} \label{regionIApp}
The first region $(x_0^+, x_v^+)$ contains the object $\mathcal{S}^{(2)}$, which describes the momentum broadening of the gluon and has a well known result in the literature. Let us review it:\\
\begin{equation}
    \mathcal{S}^{(2)} = \int_{\boldsymbol{r}_1(x_0^+)=\boldsymbol{x_g}}^{\boldsymbol{r}_1(x_v^+) = \boldsymbol{x_v}-\boldsymbol{x}}\mathcal{D}\boldsymbol{r}_1 \int_{\boldsymbol{r}_2 (x_0^+)=\boldsymbol{\overline{x}_g}}^{\boldsymbol{r}_2(x_v^+) = \boldsymbol{z_1}}\mathcal{D}\boldsymbol{r}_2 \exp{\left\{i\frac{q_0^+}{2}\int_{x_0^+}^{x_v^+}ds^+\left(\dot{\boldsymbol{r}}_1^2-\dot{\boldsymbol{r}}_2^2\right)\right\}}\mathcal{C}^{(2)} \, .
\end{equation}
Following Eq.~\eqref{color_regions} and, e.g., \cite{Apolinario:2014csa}, $\mathcal{C}^{(2)}$ reads
\begin{align}\label{C2_result}
    \mathcal{C}^{(2)} & = \frac{1}{N_c^2 -1}\left\langle \text{Tr}\left[U_A(\boldsymbol{r}_1)U_A^{\dagger}(\r_2)\right]\right\rangle = \Exp{-\frac{C_A}{2}\int ds^+ \, n(s^+)\sigma(\boldsymbol{r}_1-\boldsymbol{r}_2)} \, .
\end{align}

By changing variables to $\boldsymbol{u} = \boldsymbol{r}_1-\boldsymbol{r}_2$, $\boldsymbol{v} = \boldsymbol{r}_1+\boldsymbol{r}_2$, we end up with
\begin{equation}\label{broadening_int}
\begin{split}
    \mathcal{S}^{(2)} & = \int\mathcal{D}\boldsymbol{u}\int\mathcal{D}\boldsymbol{v}\Exp{\int ds^+\left(\frac{iq_0^+}{2}\dot{\boldsymbol{u}}\cdot\dot{\boldsymbol{v}}-\frac{C_A}{2} n(s^+)\sigma(\boldsymbol{u})\right)}\\
    & =\left(\frac{q_0^+}{2\pi\Delta_0^+}\right)^2\Exp{\frac{iq_0^+}{2\Delta_0^+}\left((\boldsymbol{x_v}-\boldsymbol{x}-\boldsymbol{x_g})^2-(\boldsymbol{z_1}-\boldsymbol{\overline{x}_g})^2\right) -\frac{C_A}{2}\int ds^+ \, n(s^+)\sigma(\boldsymbol{u}^0)} \, ,
    \end{split}
\end{equation}
where $\boldsymbol{u}^0$ is restricted to the path of a straight line due to the integral in $\boldsymbol{v}$
\begin{equation}
    \boldsymbol{u}^0(s^+) = \left((s^+-x_0^+)(\boldsymbol{x_v}-\boldsymbol{x}-\boldsymbol{z_1})-(s^+-x_v^+)(\boldsymbol{x_g}-\boldsymbol{\overline{x}_g})\right)/\Delta_0^+ \, ,
\end{equation}
and $\Delta_0^+ = x_v^+-x_0^+$. Because the remaining of the squared amplitude does not depend on the external transverse coordinates $\boldsymbol{x_g}$ and $\boldsymbol{\overline{x}_g}$, one can explicitly carry out these integrals. We start by changing variables to
\begin{equation}\label{gluon_broadening}
    \begin{split}
        & \boldsymbol{\Delta_g} = \boldsymbol{x_g} -\boldsymbol{\overline{x}_g} \, , \quad  \boldsymbol{\Sigma_g} = \boldsymbol{x_g} + \boldsymbol{\overline{x}_g}\, ,
    \end{split}
\end{equation}
which is a transformation of jacobian $1/4$. Picking up the phases that depend on $\boldsymbol{\Delta_g}$ and $\boldsymbol{\Sigma_g}$ in Eq.~\eqref{T_expanded} and realizing $\sigma(\boldsymbol{u}^0)$ only depends on $\boldsymbol{\Delta_g}$:
\begin{equation}\label{regionI_finalRes}
\begin{split}
    & \int_{\boldsymbol{x_g},\boldsymbol{\overline{x}_g}} e^{i\boldsymbol{p}\cdot\boldsymbol{x_g}}e^{-i\bar \p \cdot \overline{x}_g}\mathcal{S}^{(2)} = \frac{1}{4}\left(\frac{q_0^+}{2\pi\Delta_0^+}\right)^2 \int_{\boldsymbol{\Delta_g},\boldsymbol{\Sigma_g}} e^{i\boldsymbol{\Delta_g}\cdot\boldsymbol{\Sigma_p}/2}e^{i\boldsymbol{\Sigma_g}\cdot\boldsymbol{\Delta_p}/2}\\
    & \times \Exp{\frac{iq_0^+}{2\Delta_0^+}\Big((\boldsymbol{x_v}-\boldsymbol{x})^2-\boldsymbol{z_1}^2-\boldsymbol{\Delta_g}\cdot(\boldsymbol{x_v}-\boldsymbol{x}+\boldsymbol{z_1})-\boldsymbol{\Sigma_g}\cdot(\boldsymbol{x_v}-\boldsymbol{x}-\boldsymbol{z_1}-\boldsymbol{\Delta_g})\Big)}\\
    & \times \Exp{-\frac{C_A}{2}\int ds^+ \, n(s^+)\sigma(\boldsymbol{u}^0)}\\
    & = e^{i\boldsymbol{\Sigma_p}\cdot(\x_v-\x-\z_1-\Delta_0^+(\boldsymbol{\Delta_p}/q_0^+))/2} \,  e^{i\boldsymbol{\Delta_p}\cdot(\boldsymbol{x_v}-\x+\boldsymbol{z_1})/2}\\
    & \times \Exp{-\frac{C_A}{2}\int ds^+ \, n(s^+)\sigma(\boldsymbol{x_v}-\boldsymbol{x}-\boldsymbol{z_1} + (s^+-x_v^+)\frac{\boldsymbol{\Delta_p}}{q_0^+})}\, ,
\end{split}
\end{equation}
where we defined the placeholders $\boldsymbol{\Sigma_p} = \p + \bar\p$ and $\boldsymbol{\Delta_p} = \p - \bar\p$.

The last region $(\bar x_v^+, L^+)$ contains the object $\mathcal{S}^{(4)}$, which in the large-$N_c$ limit simplifies to the independent broadening of the quark and anti-quark (see Appendix~\ref{app:2:color_struct}) and thus to the product of two 2-point functions. It then reads 
\begin{equation}
    \mathcal{S}^{(4)} = \mathcal{S}^{(2)}_q\mathcal{S}^{(2)}_{\bar q} + \mathcal{O}\left(\frac{1}{N_c}\right)\, , 
\end{equation}
where
\begin{align}
        & \mathcal{S}^{(2)}_q = \int_{\boldsymbol{r}_1(\bar x_v^+)=\boldsymbol{z_3}}^{\boldsymbol{r}_1(L^+) = \boldsymbol{x_q}}\mathcal{D}\boldsymbol{r}_1 \int_{\boldsymbol{r}_2 (\bar x_v^+)=\boldsymbol{\bar x_v - \bar y}}^{\boldsymbol{r}_2(L^+) = \boldsymbol{\bar x_q}}\mathcal{D}\boldsymbol{r}_2 \exp{\left\{i\frac{q_0^+}{2}\int_{x_0^+}^{x_v^+}ds^+\left(\dot{\boldsymbol{r}}_1^2-\dot{\boldsymbol{r}}_2^2\right)\right\}}\mathcal{C}^{(2)}_q\nn
        & \mathcal{S}^{(2)}_{\bar q} = \int_{\boldsymbol{r}_1(\bar x_v^+)=\boldsymbol{z_2}}^{\boldsymbol{r}_1(L^+) = \boldsymbol{x_{\bar q}}}\mathcal{D}\boldsymbol{r}_1 \int_{\boldsymbol{r}_2 (\bar x_v^+)=\boldsymbol{\bar x_v}}^{\boldsymbol{r}_2(L^+) = \boldsymbol{\bar x_{\bar q}}}\mathcal{D}\boldsymbol{r}_2 \exp{\left\{i\frac{q_0^+}{2}\int_{x_0^+}^{x_v^+}ds^+\left(\dot{\boldsymbol{r}}_1^2-\dot{\boldsymbol{r}}_2^2\right)\right\}}\mathcal{C}^{(2)}_{\bar q}\, ,
\end{align}
and
\begin{align}
    & \mathcal{C}^{(2)}_q = \frac{1}{N_c}\left\langle\text{Tr}\left[U(\r_1)U^{\dagger}(\r_2)\right]\right\rangle = \Exp{-\frac{C_F}{2}\int ds^+ \, n(s^+)\sigma(\boldsymbol{r}_1-\boldsymbol{r}_2)}\nn
    & \mathcal{C}^{(2)}_{\bar q} = \frac{1}{N_c}\left\langle\text{Tr}\left[U(\r_2)U^{\dagger}(\r_1)\right]\right\rangle = \Exp{-\frac{C_F}{2}\int ds^+ \, n(s^+)\sigma(\boldsymbol{r}_2-\boldsymbol{r}_1)} \, .
\end{align}

Picking up the phases in Eq.~\eqref{T_expanded} attached to $\p_1$ and $\p_2$ and carrying out the calculation similarly to what was done for the gluon broadening $\mathcal{S}^{2}$ but with the same momenta in the direct and complex conjugate amplitudes, we get
\begin{equation}\label{regionIII_finalRes}
\begin{split}
    & \int_{\boldsymbol{x_{\bar q}},\boldsymbol{\overline{x}_{\bar q}},\boldsymbol{x_q},\boldsymbol{\overline{x}_q}} e^{-i\boldsymbol{p_1}\cdot(\boldsymbol{x_q}-\boldsymbol{\overline{x}_q})}e^{-i\boldsymbol{p_2}\cdot(\boldsymbol{x_{\bar q}}-\boldsymbol{\overline{x}_{\bar q}})} \mathcal{S}^{(4)} = e^{-i\boldsymbol{p_1}\cdot(\boldsymbol{z_3}-\boldsymbol{\overline{x}_v}+\boldsymbol{\overline{y}})} e^{-i\boldsymbol{p_2}\cdot(\boldsymbol{z_2}-\boldsymbol{\overline{x}_v})}\\
    & \times \Exp{-\frac{C_F}{2}\int_{\overline{x}_v^+}^{L^+} ds^+n(s^+)(\sigma(\boldsymbol{z_3}-\boldsymbol{\overline{x}_v}+\boldsymbol{\overline{y}})+\sigma(\boldsymbol{\overline{x}_v}-\boldsymbol{z_2}))} +\mathcal{O}\left(\frac{1}{N_c}\right)\, .
\end{split}
\end{equation}

\subsection{3-point function}\label{regionIIIApp}

The middle region $(x_v^+, \bar x_v^+)$ contains the object $\mathcal{S}^{(3)}$ describing the production of the $q\bar q$ antenna inside the medium. After changing the intermediate transverse variables from $\{\z_1,\z_2,\z_3\}$ to
\begin{equation}\label{zi_to_vi}
    \begin{split}
        & \boldsymbol{z_1}\rightarrow \boldsymbol{v_1} = \boldsymbol{x_v-\x-z_1}\\
        & \boldsymbol{z_2}\rightarrow \boldsymbol{v_2} = \boldsymbol{z_2-\overline{x}_v}\\
        & \boldsymbol{z_3}\rightarrow \boldsymbol{v_3} = \boldsymbol{z_3-\overline{x}_v+\overline{\y}}\, ,
    \end{split}
\end{equation}
we write the three body average as
\begin{align}
    \mathcal{S}^{(3)} & = \int_{\boldsymbol{r_0^i}}^{\boldsymbol{r_0^f}}\mathcal{D}\boldsymbol{r}_0 \int_{\boldsymbol{r_1^i}}^{\boldsymbol{r_1^f}}\mathcal{D}\boldsymbol{r}_1 \int_{\boldsymbol{r_2^i}}^{\boldsymbol{r_2^f}}\mathcal{D}\boldsymbol{r}_2 \Exp{\frac{i}{2}\int_{x_v^+}^{\overline{x}_v^+} ds^+\left(-q_0^+\dot{\boldsymbol{r}}_0^2+p_1^+\dot{\boldsymbol{r}}_1^2 + p_2^+\dot{\boldsymbol{r}}_2^2\right)} \mathcal{C}^{(3)}\, ,
\end{align}
where
\begin{align}
    \mathcal{C}^{(3)} = \frac{1}{2(N_c^2-1)}\left\langle \text{Tr}\left[U^{\dagger}(\r_8)U(\r_3)\right]\text{Tr}\left[U^{\dagger}(\r_4)U(\r_8)\right]  - \frac{1}{N_c}\text{Tr}\left[U^{\dagger}(\r_4)U(\r_3)\right]\right\rangle
\end{align}
and
\begin{equation}
    \begin{split}
        & \boldsymbol{r_0^i} = \boldsymbol{x_v-x-v_1} \\
        & \boldsymbol{r_1^i} = \boldsymbol{x_v-y}\\
        & \boldsymbol{r_2^i} = \boldsymbol{x_v}
    \end{split}
    \qquad \qquad
    \begin{split}
        & \boldsymbol{r_0^f} = \boldsymbol{\overline{x}_v-\overline{x}} \\
        & \boldsymbol{r_1^f} = \boldsymbol{v_3+\overline{x}_v-\overline{y}} \\
        & \boldsymbol{r_2^f} = \boldsymbol{v_2+\overline{x}_v} \, .
    \end{split}
\end{equation}
The explicit form of $\mathcal{C}^{(3)}$ is known in the literature (cf. \cite{Isaksen:2020npj, Apolinario:2014csa}). Hence, we have:
\begin{equation}\label{3point_corr}
    \begin{split}
        \mathcal{S}^{(3)} & = \frac{1}{2}\int\mathcal{D}\boldsymbol{r}_0 \int\mathcal{D}\boldsymbol{r}_1 \int\mathcal{D}\boldsymbol{r}_2 \Exp{\frac{i}{2}\int ds^+\left(-q_0^+\dot{\boldsymbol{r}}_0^2+p_1^+\dot{\boldsymbol{r}}_1^2 + p_2^+\dot{\boldsymbol{r}}_2^2\right)}\\
        & \times \Exp{-\frac{1}{2}\int ds^+ \, n(s^+)(N_c (\sigma(\boldsymbol{r}_2-\boldsymbol{r}_0)+\sigma(\boldsymbol{r}_1-\boldsymbol{r}_0))-\frac{1}{N_c}\sigma(\boldsymbol{r}_1-\boldsymbol{r}_2))}\, .
    \end{split}
\end{equation}
To further simplify these path integrals, we can make the following change of variables:
\begin{equation}
    \boldsymbol{r}_1\rightarrow \boldsymbol{u}_1 = \boldsymbol{r}_2-\boldsymbol{r}_1\, ,\quad  \boldsymbol{r}_2\rightarrow \boldsymbol{u}_2 = -\boldsymbol{r}_0+z\boldsymbol{r}_1+(1-z)\boldsymbol{r}_2 \\
\end{equation}
where $z = p_1^+/q_0^+$ as stated already. This leaves us with no kinetic term and no $\sigma$ term on $\boldsymbol{r}_0$, allowing to perform the integrals in $\boldsymbol{r}_0$ and $\boldsymbol{u}_2$, restricting the latter to a straight line path (much like the integration in Eq.~\eqref{gluon_broadening}):
\begin{equation}
    \begin{split}
         \mathcal{S}^{(3)} & = \frac{1}{2}\int\mathcal{D}\boldsymbol{r}_0\int\mathcal{D}\boldsymbol{u}_1\int\mathcal{D}\boldsymbol{u}_2 \Exp{\frac{iq_0^+}{2}\int ds^+\left(2\dot{\boldsymbol{r}}_0\cdot \dot{\boldsymbol{u}}_2+\dot{\boldsymbol{u}}_2^2+z(1-z)\dot{\boldsymbol{u}}_1^2\right)}\\
         & \times \Exp{-\frac{1}{2}\int ds^+ \, n(s^+)(N_c (\sigma(\boldsymbol{u}_2+z\boldsymbol{u}_1)+\sigma(\boldsymbol{u}_2-(1-z)\boldsymbol{u}_1))-\frac{1}{N_c}\sigma(-\boldsymbol{u}_1))}\\
        & = \frac{1}{2}\left(\frac{q_0^+}{2\pi\Delta t}\right)^2 \Exp{\frac{iq_0^+}{2}\frac{\Delta\boldsymbol{u}_2}{\Delta t}\left(2\Delta\boldsymbol{r}_0 + \Delta\boldsymbol{u}_2\right)}\cK(\boldsymbol{u}_1^i, \boldsymbol{u}_1^f, \boldsymbol{u}_2^i,\boldsymbol{u}_2^f)\, ,
    \end{split}
\end{equation}
with $\tilde{q}_0^+ = z(1-z)q_0^+$, $\Delta t = \overline{x}_v^+ - x_v^+$ and
\begin{equation*}
\begin{split}
    \boldsymbol{u}_2^0(s^+) = ((s^+-x_v^+)\boldsymbol{u}_2^f-(s^+-\overline{x}_v^+)\boldsymbol{u}_2^i)/\Delta t\, ,\\
    \end{split}
\end{equation*}
\begin{equation}
\begin{split}
        & \boldsymbol{r_0^i} = \boldsymbol{x_v-x-v_1} \\
        & \boldsymbol{u_1^i} = \boldsymbol{y}\\
        & \boldsymbol{u_2^i} = \boldsymbol{v_1}+\x-z\boldsymbol{y}
    \end{split}
    \qquad \qquad
    \begin{split}
        & \boldsymbol{r_0^f} = \boldsymbol{\overline{x}_v-\overline{x}}\\
        & \boldsymbol{u_1^f} = \boldsymbol{v_2}-\boldsymbol{v_3}+\boldsymbol{\overline{y}} \\
        & \boldsymbol{u_2^f} = \boldsymbol{v_2}+\overline{\x}-z(\boldsymbol{v_2}-\boldsymbol{v_3}+\boldsymbol{\overline{y}})\, .\\
\end{split}
\end{equation}
We have also defined the emission \textit{kernel} $\cK$ as
\begin{align}
    \cK(\boldsymbol{u}_1^i,& \boldsymbol{u}_1^f, \boldsymbol{u}_2^i,\boldsymbol{u}_2^f) \equiv \int\mathcal{D}\boldsymbol{u}_1\Exp{\frac{i\tilde{q}_0^+}{2}\int ds^+\dot{\boldsymbol{u}}_1^2}\nn
    & \times \Exp{-\frac{1}{2}\int ds^+ \, n(s^+)(N_c (\sigma(\boldsymbol{u}_2^0+z\boldsymbol{u}_1)+\sigma(\boldsymbol{u}_2^0-(1-z)\boldsymbol{u}_1))-\frac{1}{N_c}\sigma(-\boldsymbol{u}_1))} \, .
\end{align}
It is interesting to note that $\mathcal{S}^{(3)}$ only depends on $\x_v$ and $\bar\x_v$ through the difference $\Delta_v = \bar\x_v-\x_v$ in $\boldsymbol{\Delta r}_0$. Aside from this dependence, after applying the change of variables in Eq.~\eqref{zi_to_vi} to the results in Eq.~\eqref{regionI_finalRes} and \eqref{regionIII_finalRes}, we see that the only dependence on the vertex transverse positions that is left is via a complex phase $e^{i(\p -\bar \p)\cdot\x_v}$ in Eq.~\eqref{regionI_finalRes}. This allows one to integrate over both $\x_v$ and $\bar \p$ to set $\p = \bar\p$, i.e., the same transverse momentum flows out of the initial current in amplitude and its complex conjugate. This leads to the result in Eq.~\eqref{eq:dN_1}.

\bibliographystyle{jhep}
\bibliography{refs.bib}

\providecommand{\href}[2]{#2}\begingroup\raggedright\begin{thebibliography}{10}

\bibitem{PHENIX:2001hpc}
{\bf PHENIX} Collaboration, K.~Adcox et~al., {\it {Suppression of hadrons with large transverse momentum in central Au+Au collisions at $\sqrt{s_{NN}}$ = 130-GeV}},  {\em Phys. Rev. Lett.} {\bf 88} (2002) 022301, [\href{http://arxiv.org/abs/nucl-ex/0109003}{{\tt nucl-ex/0109003}}].

\bibitem{STAR:2005gfr}
{\bf STAR} Collaboration, J.~Adams et~al., {\it {Experimental and theoretical challenges in the search for the quark gluon plasma: The STAR Collaboration's critical assessment of the evidence from RHIC collisions}},  {\em Nucl. Phys. A} {\bf 757} (2005) 102--183, [\href{http://arxiv.org/abs/nucl-ex/0501009}{{\tt nucl-ex/0501009}}].

\bibitem{STAR:2020xiv}
{\bf STAR} Collaboration, J.~Adam et~al., {\it {Measurement of inclusive charged-particle jet production in Au + Au collisions at $\sqrt{s_{NN}}=$200 GeV}},  {\em Phys. Rev. C} {\bf 102} (2020), no.~5 054913, [\href{http://arxiv.org/abs/2006.00582}{{\tt arXiv:2006.00582}}].

\bibitem{Muller:2012zq}
B.~Muller, J.~Schukraft, and B.~Wyslouch, {\it {First Results from Pb+Pb collisions at the LHC}},  {\em Ann. Rev. Nucl. Part. Sci.} {\bf 62} (2012) 361--386, [\href{http://arxiv.org/abs/1202.3233}{{\tt arXiv:1202.3233}}].

\bibitem{ATLAS:2018gwx}
{\bf ATLAS} Collaboration, M.~Aaboud et~al., {\it {Measurement of the nuclear modification factor for inclusive jets in Pb+Pb collisions at $\sqrt{s_\mathrm{NN}}=5.02$ TeV with the ATLAS detector}},  {\em Phys. Lett. B} {\bf 790} (2019) 108--128, [\href{http://arxiv.org/abs/1805.05635}{{\tt arXiv:1805.05635}}].

\bibitem{CMS:2021vui}
{\bf CMS} Collaboration, A.~M. Sirunyan et~al., {\it {First measurement of large area jet transverse momentum spectra in heavy-ion collisions}},  {\em JHEP} {\bf 05} (2021) 284, [\href{http://arxiv.org/abs/2102.13080}{{\tt arXiv:2102.13080}}].

\bibitem{Busza:2018rrf}
W.~Busza, K.~Rajagopal, and W.~van~der Schee, {\it {Heavy Ion Collisions: The Big Picture, and the Big Questions}},  {\em Ann. Rev. Nucl. Part. Sci.} {\bf 68} (2018) 339--376, [\href{http://arxiv.org/abs/1802.04801}{{\tt arXiv:1802.04801}}].

\bibitem{Schlichting:2019abc}
S.~Schlichting and D.~Teaney, {\it {The First fm/c of Heavy-Ion Collisions}},  {\em Ann. Rev. Nucl. Part. Sci.} {\bf 69} (2019) 447--476, [\href{http://arxiv.org/abs/1908.02113}{{\tt arXiv:1908.02113}}].

\bibitem{Berges:2020fwq}
J.~Berges, M.~P. Heller, A.~Mazeliauskas, and R.~Venugopalan, {\it {QCD thermalization: Ab initio approaches and interdisciplinary connections}},  {\em Rev. Mod. Phys.} {\bf 93} (2021), no.~3 035003, [\href{http://arxiv.org/abs/2005.12299}{{\tt arXiv:2005.12299}}].

\bibitem{Jaiswal:2016hex}
A.~Jaiswal and V.~Roy, {\it {Relativistic hydrodynamics in heavy-ion collisions: general aspects and recent developments}},  {\em Adv. High Energy Phys.} {\bf 2016} (2016) 9623034, [\href{http://arxiv.org/abs/1605.08694}{{\tt arXiv:1605.08694}}].

\bibitem{Qin:2015srf}
G.-Y. Qin and X.-N. Wang, {\it {Jet quenching in high-energy heavy-ion collisions}},  {\em Int. J. Mod. Phys. E} {\bf 24} (2015), no.~11 1530014, [\href{http://arxiv.org/abs/1511.00790}{{\tt arXiv:1511.00790}}].

\bibitem{Cunqueiro:2021wls}
L.~Cunqueiro and A.~M. Sickles, {\it {Studying the QGP with Jets at the LHC and RHIC}},  {\em Prog. Part. Nucl. Phys.} {\bf 124} (2022) 103940, [\href{http://arxiv.org/abs/2110.14490}{{\tt arXiv:2110.14490}}].

\bibitem{Apolinario:2022vzg}
L.~Apolin\'ario, Y.-J. Lee, and M.~Winn, {\it {Heavy quarks and jets as probes of the QGP}},  {\em Prog. Part. Nucl. Phys.} {\bf 127} (2022) 103990, [\href{http://arxiv.org/abs/2203.16352}{{\tt arXiv:2203.16352}}].

\bibitem{Wang:2002ri}
E.~Wang and X.-N. Wang, {\it {Jet tomography of dense and nuclear matter}},  {\em Phys. Rev. Lett.} {\bf 89} (2002) 162301, [\href{http://arxiv.org/abs/hep-ph/0202105}{{\tt hep-ph/0202105}}].

\bibitem{Vitev:2004bh}
I.~Vitev, {\it {Jet tomography}},  {\em J. Phys. G} {\bf 30} (2004) S791--S800, [\href{http://arxiv.org/abs/hep-ph/0403089}{{\tt hep-ph/0403089}}].

\bibitem{Armesto:2004pt}
N.~Armesto, C.~A. Salgado, and U.~A. Wiedemann, {\it {Measuring the collective flow with jets}},  {\em Phys. Rev. Lett.} {\bf 93} (2004) 242301, [\href{http://arxiv.org/abs/hep-ph/0405301}{{\tt hep-ph/0405301}}].

\bibitem{Hauksson:2023tze}
S.~Hauksson and E.~Iancu, {\it {Jet polarisation in an anisotropic medium}},  {\em JHEP} {\bf 08} (2023) 027, [\href{http://arxiv.org/abs/2303.03914}{{\tt arXiv:2303.03914}}].

\bibitem{Baty:2021ugw}
A.~Baty, P.~Gardner, and W.~Li, {\it {Novel observables for exploring QCD collective evolution and quantum entanglement within individual jets}},  {\em Phys. Rev. C} {\bf 107} (2023), no.~6 064908, [\href{http://arxiv.org/abs/2104.11735}{{\tt arXiv:2104.11735}}].

\bibitem{Zhao:2024wqs}
W.~Zhao, Z.-W. Lin, and X.-N. Wang, {\it {Collectivity inside high-multiplicity jets in high-energy proton-proton collisions}},  \href{http://arxiv.org/abs/2401.13137}{{\tt arXiv:2401.13137}}.

\bibitem{Boguslavski:2023waw}
K.~Boguslavski, A.~Kurkela, T.~Lappi, F.~Lindenbauer, and J.~Peuron, {\it {Jet quenching parameter in QCD kinetic theory}},  \href{http://arxiv.org/abs/2312.00447}{{\tt arXiv:2312.00447}}.

\bibitem{Boguslavski:2023alu}
K.~Boguslavski, A.~Kurkela, T.~Lappi, F.~Lindenbauer, and J.~Peuron, {\it {Jet momentum broadening during initial stages in heavy-ion collisions}},  {\em Phys. Lett. B} {\bf 850} (2024) 138525, [\href{http://arxiv.org/abs/2303.12595}{{\tt arXiv:2303.12595}}].

\bibitem{Ipp:2020mjc}
A.~Ipp, D.~I. M\"uller, and D.~Schuh, {\it {Anisotropic momentum broadening in the 2+1D Glasma: analytic weak field approximation and lattice simulations}},  {\em Phys. Rev. D} {\bf 102} (2020), no.~7 074001, [\href{http://arxiv.org/abs/2001.10001}{{\tt arXiv:2001.10001}}].

\bibitem{Ipp:2020nfu}
A.~Ipp, D.~I. M\"uller, and D.~Schuh, {\it {Jet momentum broadening in the pre-equilibrium Glasma}},  {\em Phys. Lett. B} {\bf 810} (2020) 135810, [\href{http://arxiv.org/abs/2009.14206}{{\tt arXiv:2009.14206}}].

\bibitem{Avramescu:2023qvv}
D.~Avramescu, V.~B\u{a}ran, V.~Greco, A.~Ipp, D.~I. M\"uller, and M.~Ruggieri, {\it {Simulating jets and heavy quarks in the glasma using the colored particle-in-cell method}},  {\em Phys. Rev. D} {\bf 107} (2023), no.~11 114021, [\href{http://arxiv.org/abs/2303.05599}{{\tt arXiv:2303.05599}}].

\bibitem{Carrington:2021dvw}
M.~E. Carrington, A.~Czajka, and S.~Mrowczynski, {\it {Jet quenching in glasma}},  {\em Phys. Lett. B} {\bf 834} (2022) 137464, [\href{http://arxiv.org/abs/2112.06812}{{\tt arXiv:2112.06812}}].

\bibitem{Carrington:2022bnv}
M.~E. Carrington, A.~Czajka, and S.~Mrowczynski, {\it {Transport of hard probes through glasma}},  {\em Phys. Rev. C} {\bf 105} (2022), no.~6 064910, [\href{http://arxiv.org/abs/2202.00357}{{\tt arXiv:2202.00357}}].

\bibitem{Barata:2024xwy}
J.~Barata, S.~Hauksson, X.~Mayo~L\'opez, and A.~V. Sadofyev, {\it {Jet quenching in the glasma phase: medium-induced radiation}},  \href{http://arxiv.org/abs/2406.07615}{{\tt arXiv:2406.07615}}.

\bibitem{Kuzmin:2023hko}
M.~V. Kuzmin, X.~Mayo~L\'opez, J.~Reiten, and A.~V. Sadofyev, {\it {Jet quenching in anisotropic flowing matter}},  {\em Phys. Rev. D} {\bf 109} (2024), no.~1 014036, [\href{http://arxiv.org/abs/2309.00683}{{\tt arXiv:2309.00683}}].

\bibitem{Barata:2023zqg}
J.~Barata, J.~G. Milhano, and A.~V. Sadofyev, {\it {Picturing QCD jets in anisotropic matter: from jet shapes to energy energy correlators}},  {\em Eur. Phys. J. C} {\bf 84} (2024), no.~2 174, [\href{http://arxiv.org/abs/2308.01294}{{\tt arXiv:2308.01294}}].

\bibitem{Barata:2023qds}
J.~Barata, X.~Mayo~L\'opez, A.~V. Sadofyev, and C.~A. Salgado, {\it {Medium induced gluon spectrum in dense inhomogeneous matter}},  {\em Phys. Rev. D} {\bf 108} (2023), no.~3 034018, [\href{http://arxiv.org/abs/2304.03712}{{\tt arXiv:2304.03712}}].

\bibitem{Barata:2022utc}
J.~Barata, A.~V. Sadofyev, and X.-N. Wang, {\it {Quantum partonic transport in QCD matter}},  {\em Phys. Rev. D} {\bf 107} (2023), no.~5 L051503, [\href{http://arxiv.org/abs/2210.06519}{{\tt arXiv:2210.06519}}].

\bibitem{Andres:2022ndd}
C.~Andres, F.~Dominguez, A.~V. Sadofyev, and C.~A. Salgado, {\it {Jet broadening in flowing matter: Resummation}},  {\em Phys. Rev. D} {\bf 106} (2022), no.~7 074023, [\href{http://arxiv.org/abs/2207.07141}{{\tt arXiv:2207.07141}}].

\bibitem{Barata:2022krd}
J.~Barata, A.~V. Sadofyev, and C.~A. Salgado, {\it {Jet broadening in dense inhomogeneous matter}},  {\em Phys. Rev. D} {\bf 105} (2022), no.~11 114010, [\href{http://arxiv.org/abs/2202.08847}{{\tt arXiv:2202.08847}}].

\bibitem{Sadofyev:2021ohn}
A.~V. Sadofyev, M.~D. Sievert, and I.~Vitev, {\it {Ab~initio coupling of jets to collective flow in the opacity expansion approach}},  {\em Phys. Rev. D} {\bf 104} (2021), no.~9 094044, [\href{http://arxiv.org/abs/2104.09513}{{\tt arXiv:2104.09513}}].

\bibitem{He:2020iow}
Y.~He, L.-G. Pang, and X.-N. Wang, {\it {Gradient Tomography of Jet Quenching in Heavy-Ion Collisions}},  {\em Phys. Rev. Lett.} {\bf 125} (2020), no.~12 122301, [\href{http://arxiv.org/abs/2001.08273}{{\tt arXiv:2001.08273}}].

\bibitem{Xiao:2024ffk}
Y.-X. Xiao, Y.~He, L.-G. Pang, H.~Zhang, and X.-N. Wang, {\it {Asymmetric jet shapes with 2D jet tomography}},  \href{http://arxiv.org/abs/2402.00264}{{\tt arXiv:2402.00264}}.

\bibitem{Fu:2022idl}
Y.~Fu, J.~Casalderrey-Solana, and X.-N. Wang, {\it {Asymmetric transverse momentum broadening in an inhomogeneous medium}},  {\em Phys. Rev. D} {\bf 107} (2023), no.~5 054038, [\href{http://arxiv.org/abs/2204.05323}{{\tt arXiv:2204.05323}}].

\bibitem{He:2022evt}
Y.~He, W.~Chen, T.~Luo, S.~Cao, L.-G. Pang, and X.-N. Wang, {\it {Event-by-event jet anisotropy and hard-soft tomography of the quark-gluon plasma}},  {\em Phys. Rev. C} {\bf 106} (2022), no.~4 044904, [\href{http://arxiv.org/abs/2201.08408}{{\tt arXiv:2201.08408}}].

\bibitem{Kuzmin:2024smy}
M.~V. Kuzmin and X.~Mayo~L\'opez, {\it {Gluon radiation inside a flowing medium}},  \href{http://arxiv.org/abs/2406.14628}{{\tt arXiv:2406.14628}}.

\bibitem{Hauksson:2020wsm}
S.~Hauksson, S.~Jeon, and C.~Gale, {\it {Probes of the quark-gluon plasma and plasma instabilities}},  {\em Phys. Rev. C} {\bf 103} (2021) 064904, [\href{http://arxiv.org/abs/2012.03640}{{\tt arXiv:2012.03640}}].

\bibitem{Hauksson:2021okc}
S.~Hauksson, S.~Jeon, and C.~Gale, {\it {Momentum broadening of energetic partons in an anisotropic plasma}},  {\em Phys. Rev. C} {\bf 105} (2022), no.~1 014914, [\href{http://arxiv.org/abs/2109.04575}{{\tt arXiv:2109.04575}}].

\bibitem{Zakharov:1996fv}
B.~G. Zakharov, {\it {Fully quantum treatment of the Landau-Pomeranchuk-Migdal effect in QED and QCD}},  {\em JETP Lett.} {\bf 63} (1996) 952--957, [\href{http://arxiv.org/abs/hep-ph/9607440}{{\tt hep-ph/9607440}}].

\bibitem{Baier:1996kr}
R.~Baier, Y.~L. Dokshitzer, A.~H. Mueller, S.~Peigne, and D.~Schiff, {\it {Radiative energy loss of high-energy quarks and gluons in a finite volume quark - gluon plasma}},  {\em Nucl. Phys. B} {\bf 483} (1997) 291--320, [\href{http://arxiv.org/abs/hep-ph/9607355}{{\tt hep-ph/9607355}}].

\bibitem{Attems:2022ubu}
M.~Attems, J.~Brewer, G.~M. Innocenti, A.~Mazeliauskas, S.~Park, W.~van~der Schee, and U.~A. Wiedemann, {\it {The medium-modified $ g\to c\overline{c} $ splitting function in the BDMPS-Z formalism}},  {\em JHEP} {\bf 01} (2023) 080, [\href{http://arxiv.org/abs/2203.11241}{{\tt arXiv:2203.11241}}].

\bibitem{Mehtar-Tani:2013pia}
Y.~Mehtar-Tani, J.~G. Milhano, and K.~Tywoniuk, {\it {Jet physics in heavy-ion collisions}},  {\em Int. J. Mod. Phys. A} {\bf 28} (2013) 1340013, [\href{http://arxiv.org/abs/1302.2579}{{\tt arXiv:1302.2579}}].

\bibitem{Blaizot:2015lma}
J.-P. Blaizot and Y.~Mehtar-Tani, {\it {Jet Structure in Heavy Ion Collisions}},  {\em Int. J. Mod. Phys. E} {\bf 24} (2015), no.~11 1530012, [\href{http://arxiv.org/abs/1503.05958}{{\tt arXiv:1503.05958}}].

\bibitem{Blaizot:2012fh}
J.-P. Blaizot, F.~Dominguez, E.~Iancu, and Y.~Mehtar-Tani, {\it {Medium-induced gluon branching}},  {\em JHEP} {\bf 01} (2013) 143, [\href{http://arxiv.org/abs/1209.4585}{{\tt arXiv:1209.4585}}].

\bibitem{Barata:2023uoi}
J.~Barata, J.-P. Blaizot, and Y.~Mehtar-Tani, {\it {Quantum to classical parton dynamics in QCD media}},  {\em Phys. Rev. D} {\bf 108} (2023), no.~1 014039, [\href{http://arxiv.org/abs/2305.10476}{{\tt arXiv:2305.10476}}].

\bibitem{Blaizot:2017ypk}
J.-P. Blaizot and M.~A. Escobedo, {\it {Quantum and classical dynamics of heavy quarks in a quark-gluon plasma}},  {\em JHEP} {\bf 06} (2018) 034, [\href{http://arxiv.org/abs/1711.10812}{{\tt arXiv:1711.10812}}].

\bibitem{Blaizot:1996az}
J.-P. Blaizot and E.~Iancu, {\it {Lifetimes of quasiparticles and collective excitations in hot QED plasmas}},  {\em Phys. Rev. D} {\bf 55} (1997) 973--996, [\href{http://arxiv.org/abs/hep-ph/9607303}{{\tt hep-ph/9607303}}].

\bibitem{Braaten:1990it}
E.~Braaten and R.~D. Pisarski, {\it {Calculation of the gluon damping rate in hot QCD}},  {\em Phys. Rev. D} {\bf 42} (1990) 2156--2160.

\bibitem{Mehtar-Tani:2019tvy}
Y.~Mehtar-Tani, {\it {Gluon bremsstrahlung in finite media beyond multiple soft scattering approximation}},  {\em JHEP} {\bf 07} (2019) 057, [\href{http://arxiv.org/abs/1903.00506}{{\tt arXiv:1903.00506}}].

\bibitem{Barata:2020sav}
J.~Barata and Y.~Mehtar-Tani, {\it {Improved opacity expansion at NNLO for medium induced gluon radiation}},  {\em JHEP} {\bf 10} (2020) 176, [\href{http://arxiv.org/abs/2004.02323}{{\tt arXiv:2004.02323}}].

\bibitem{Apolinario:2014csa}
L.~Apolin\'ario, N.~Armesto, J.~G. Milhano, and C.~A. Salgado, {\it {Medium-induced gluon radiation and colour decoherence beyond the soft approximation}},  {\em JHEP} {\bf 02} (2015) 119, [\href{http://arxiv.org/abs/1407.0599}{{\tt arXiv:1407.0599}}].

\bibitem{Kleinert:2004ev}
H.~Kleinert, {\it {Path Integrals in Quantum Mechanics, Statistics, Polymer Physics, and Financial Markets}}, .

\bibitem{Snellings:2011sz}
R.~Snellings, {\it {Elliptic Flow: A Brief Review}},  {\em New J. Phys.} {\bf 13} (2011) 055008, [\href{http://arxiv.org/abs/1102.3010}{{\tt arXiv:1102.3010}}].

\bibitem{Kharzeev:2015znc}
D.~E. Kharzeev, J.~Liao, S.~A. Voloshin, and G.~Wang, {\it {Chiral magnetic and vortical effects in high-energy nuclear collisions\textemdash{}A status report}},  {\em Prog. Part. Nucl. Phys.} {\bf 88} (2016) 1--28, [\href{http://arxiv.org/abs/1511.04050}{{\tt arXiv:1511.04050}}].

\bibitem{Kumar:2022ylt}
A.~Kumar, B.~M\"uller, and D.-L. Yang, {\it {Spin polarization and correlation of quarks from the glasma}},  {\em Phys. Rev. D} {\bf 107} (2023), no.~7 076025, [\href{http://arxiv.org/abs/2212.13354}{{\tt arXiv:2212.13354}}].

\bibitem{Du:2023izb}
X.~Du, {\it {Heavy quark drag and diffusion coefficients in the prehydrodynamic QCD plasma}},  {\em Phys. Rev. C} {\bf 109} (2024), no.~1 014901, [\href{http://arxiv.org/abs/2306.02530}{{\tt arXiv:2306.02530}}].

\bibitem{Binosi:2003yf}
D.~Binosi and L.~Theussl, {\it {JaxoDraw: A Graphical user interface for drawing Feynman diagrams}},  {\em Comput. Phys. Commun.} {\bf 161} (2004) 76--86, [\href{http://arxiv.org/abs/hep-ph/0309015}{{\tt hep-ph/0309015}}].

\bibitem{Fukushima:2007dy}
K.~Fukushima and Y.~Hidaka, {\it {Light projectile scattering off the color glass condensate}},  {\em JHEP} {\bf 06} (2007) 040, [\href{http://arxiv.org/abs/0704.2806}{{\tt arXiv:0704.2806}}].

\bibitem{Fukushima:2017mko}
K.~Fukushima and Y.~Hidaka, {\it {General formulae for dipole Wilson line correlators with the Color Glass Condensate}},  {\em JHEP} {\bf 11} (2017) 114, [\href{http://arxiv.org/abs/1708.03051}{{\tt arXiv:1708.03051}}].

\bibitem{Isaksen:2020npj}
J.~H. Isaksen and K.~Tywoniuk, {\it {Wilson line correlators beyond the large-N$_{c}$}},  {\em JHEP} {\bf 21} (2020) 125, [\href{http://arxiv.org/abs/2107.02542}{{\tt arXiv:2107.02542}}].

\bibitem{Dominguez:2012ad}
F.~Dominguez, C.~Marquet, A.~M. Stasto, and B.-W. Xiao, {\it {Universality of multiparticle production in QCD at high energies}},  {\em Phys. Rev. D} {\bf 87} (2013) 034007, [\href{http://arxiv.org/abs/1210.1141}{{\tt arXiv:1210.1141}}].

\bibitem{Dominguez:2019ges}
F.~Dom\'\i{}nguez, J.~G. Milhano, C.~A. Salgado, K.~Tywoniuk, and V.~Vila, {\it {Mapping collinear in-medium parton splittings}},  {\em Eur. Phys. J. C} {\bf 80} (2020), no.~1 11, [\href{http://arxiv.org/abs/1907.03653}{{\tt arXiv:1907.03653}}].

\end{thebibliography}\endgroup

\end{document}